\documentclass[journal]{IEEEtran}

\usepackage{tikz}
\usepackage{tikz-3dplot}
\usepackage{pgfplots}
 \pgfplotsset{compat=newest}
 
\usetikzlibrary{positioning}

\usepackage[latin1]{inputenc}
\usepackage{amsmath}
\usepackage{amsfonts}
\usepackage{amssymb}
\usepackage{graphicx}
\usepackage{color}
\usepackage{dsfont}
\usepackage{hyperref}
\usepackage{import}
\usepackage{tcolorbox}
\usepackage{subcaption}

\usepackage{amssymb,tikz,fancyhdr,enumerate,array,booktabs,setspace,pgf,tikz,fancyhdr,graphicx,color}
\usepackage{pgfplots}

\pgfplotsset{compat=1.10}
\usepgfplotslibrary{fillbetween}
\usepackage[T1]{fontenc}
\usetikzlibrary{patterns,arrows,decorations.pathreplacing}	
\usepackage{pgfplots}

\usepackage{mathcomSTEv4}

\newtheorem{lemma}{Lemma}
\newtheorem{Theorem}{Theorem}
\newtheorem{defi}{Definition}
\newtheorem{rem}{Remark}

\newtheorem{Cor}{Corollary}

\newcommand{\rank}{{\rm rank}}

\newcommand{\rvline}{\hspace*{-\arraycolsep}\vline\hspace*{-\arraycolsep}}
\renewcommand{\xiv}{\boldsymbol{\xi}}
\newcommand{\xitv}{\tilde{\boldsymbol{\xi}}}
\newcommand{\changed}{{\color{red} }}
\newcommand{\dm}{\mathrm{d}}
\newcommand{\cm}{\mathrm{c}}
\newcommand{\spanM}{{\rm span}}
\newcommand{\spark}{{\rm SPARK}}
\graphicspath{{"Figures/"}}
\allowdisplaybreaks

\usepackage[printonlyused]{acronym}

\acrodef{dcarma}[DCE-ARMA]{\emph{Discrete--Continuous Excitation Autoregressive--Moving--Average}}

\acrodef{dce}[DCE-ARMA]{\emph{Discrete--Continuous Excitation Autoregressive--Moving--Average}}

\acrodef{asrv}[ASRV]{\emph{Affinely Singular Random Vectors}}

\acrodef{rid}[RID]{\emph{R\'enyi Information Dimension}}
\acrodef{rdf}[RDF]{\emph{Rate-Distortion Function}}
\acrodef{rv}[RV]{\emph{Random Variable/Vector}}

\acrodef{rid}[RID]{\emph{R\'enyi Information Dimension}}
\acrodef{idr}[IDR]{\emph{Information Dimension Rate}}
\acrodef{bid}[BID]{\emph{Block Information Dimension}}
\acrodef{arma}[ARMA]{\emph{Auto-Regressive Moving-Average}}
\acrodef{ma}[MA]{\emph{Moving-Average}}
\acrodef{ar}[AR]{\emph{Autoregressive}}

\title{
ARMA Processes with Discrete-Continuous Excitation: Compressibility Beyond Sparsity
}

\author{{Mohammad-Amin~Charusaie},~Arash~Amini,~\IEEEmembership{Senior,~IEEE},~and~Stefano~Rini,~\IEEEmembership{Senior,~IEEE}
\thanks{M. Charusaie  was with Sharif university of technology, Tehran, Iran, at the time of writing this manuscript and is now with Max Planck Institute for Intelligent Systems, T{\"u}bingen, Germany (email: mcharusaie@tuebingen.mpg.de). }
\thanks{A. Amini is with the department of Electrical Engineering, Sharif University of Technology, Tehran, Iran (email: aamini@sharif.edu).}%
\thanks{S. Rini is with the Electrical and Computer Engineering department, National Yang-Ming Chiao-Tung University (NYCU), Hsinchu, Taiwan. (email: stefano.rini@nycu.edu.tw)}}

\definecolor{orange}{RGB}{255,127,0}
\definecolor{byzantium}{rgb}{0.44, 0.16, 0.39}
\definecolor{byzantine}{rgb}{0.74, 0.2, 0.64}

\begin{document}

\maketitle

\begin{abstract}
R\'enyi Information Dimension (RID) plays a central role in quantifying the compressibility of random variables with singularities in their distribution, encompassing and extending beyond the class of sparse sources. The RID, from a high perspective, presents the average number of bits that is needed for coding the i.i.d. samples of a random variable with high precision. 
There are two main extensions of the RID for stochastic processes: information dimension rate (IDR) and block information dimension (BID). In addition, a more recent approach towards the compressibility of stochastic processes revolves around the concept of $\ep$-achievable compression rates, which treat a random process as the limiting point of finite-dimensional random vectors and apply the compressed sensing tools on these random variables. While there is limited knowledge about the interplay of the the BID, the IDR, and $\ep$-achievable compression rates, the value of IDR and BID themselves are known only for very specific types of processes, namely  i.i.d. sequences (i.e., discrete-domain white noise) and moving-average (MA) processes.
This paper investigates the IDR and BID of discrete-time Auto-Regressive Moving-Average (ARMA) processes in general, and their relations with $\ep$-achievable compression rates when the excitation noise has a discrete-continuous measure. To elaborate, this paper shows that the RID and $\ep$-achievable compression rates of this type of processes are equal to that of their excitation noise. In other words, the samples of such ARMA processes can be compressed as much as their sparse excitation noise, although the samples themselves are by no means sparse.
The results of this paper can be used to evaluate the compressibility of various types of locally correlated data with finite- or infinite-memory as they are often modelled via ARMA processes.

\end{abstract}

\begin{IEEEkeywords}
ARMA processes, discrete-continuous random variables, 
R\'enyi Information Dimension.
\end{IEEEkeywords}

Discrete-domain auto-regressive moving-average (ARMA) processes are popular stochastic models for explaining  data with  long-range dependencies; these models are used for estimation and classification purposes \cite{Cadzow82}.  Data with long-range dependencies occur in phenomena such as network traffic \cite{parsimonious}, fading channels  \cite{armasynthesis}, fluid velocity \cite{lda}, solar irradiance  \cite{david2016probabilistic},  automotive traffic \cite{klepsch2017prediction}, and housing investments \cite{miles2009irreversibility}. 

%

These processes consist of a shaping filter that acts on an innovation process, also referred to as an excitation noise. The distribution laws of the ARMA model is determined by the innovation process, whereas the dependency patterns among samples of the ARMA process is controlled by the shaping filter, which is in turn determined by a set of parameters (AR/MA parameters). The main advantage of ARMA models is that they allow for deriving optimal estimators in certain settings. However, the  distribution of a statistical model and more specifically, its compressibility properties, is a key factor in forming an effective model for realistic applications \cite{bostan2013sparse}. 


Despite the important role of ARMA processes, their compressibility is investigated only in special cases, e.g., auto-regressive processes with Gaussian innovations \cite{Gray1970}. While Gaussian models often lend themselves to analytical solutions and closed-form expressions, they do not adequately capture a significant portion of natural signals, including spectral, seismic, and biological data. These data are better described by sparsity inducing laws such as Bernoulli-Gaussian \cite{soussen2011bernoulli}. However, conventional approaches fail to measure the compressibility of an ARMA processes with such sparsity inducing laws, and alternative approaches have yet to emerge in the literature.


In this paper, we evaluate the compressibility of ARMA processes in a rather broad setting. Our main result is to show that the information-theoretic compressibility of an ARMA process is equal to that of its innovation process, and independent of its AR/MA parameters. 
Furthermore, this value coincides with the notion of smooth and robust compressibility in \cite{wu2010renyi} which extends some compressed sensing concepts to stochastic processes.
%
Our approach results are derived using a technique that is recently developed in \cite{CharusaieISIT2020P1} and quantifies the compressibility of random vectors with affinely singular distributions. In simple words, the singular part (e.g. mass probabilities) of the distribution of these vectors are supported on affine subsets. By applying this technique, we show that  finite dimensional samples of ARMA processes with discrete-continuous excitation noise are affinely singular random vectors. In turn, this result enables us to quantify the smooth and robust compressibility of such ARMA processes.

\subsection*{Relevant Literature}

Information-theoretic compressibility of discrete-domain stochastic processes is studied mainly in special cases. In \cite{franke1985arma} it is shown that ARMA processes have the highest entropy among all processes with a given covariance matrix. In \cite{Gray1970}, it is proved that the rate-distortion function (RDF) of an autoregressive process with Gaussian excitation is equal to that of its excitation noise. A similar result is  shown for first-order autoregressive processes, i.e., known as AR(1) processes, with discrete excitation. The results of \cite{Gray1970} further implies that information dimension rates of such processes are equal to that of their excitation noise.
The RDF of non-stationary Gaussian  autoregressive processes and  vector Gaussian  autoregressive processes are studied in \cite{gray2008note}\textendash\cite{hashimoto1980rate} and  \cite{kafedziski2005rate}, respectively. The RDF of the process obtained by applying an FIR filter to a general  wide-sense stationary Gaussian process is covered in \cite{gutierrez2008asymptotically}.
For this class of processios, the authors of
\cite{gutierrez11} derive the differential entropy rate.
A learning-based compression approach is introduced in \cite{Jalali20} which achieves the block information dimension on a class of stationary processes; we note that this technique is applicable to  some MA processes studied in this paper.


Almost lossless compression rates, also referred to as $\ep$-achievable compression rates, are introduced in \cite{wu2010renyi} and studied for i.i.d. processes. In \cite{Gutman20}, fundamental bounds are derived for worst $\ep$-achievable compression-rates of bounded processes. In \cite{CharusaieISIT2020P1}, almost lossless compression rates of MA processes are investigated; the considered processes are examples of processes with affinely singular sample distributions.

Several other measures of compressibility for stochastic processes are proposed in the literature.
An energy-based measure of compressibility is introduced and studied for continuous i.i.d. random variables in \cite{amini11} and for ergodic sequences in \cite{silva15}. 
The non-asymptotic lossless compression rates of a random vector has been studied in \cite{Alberti19} via modifying the Minkowski dimension of the support set of the vector.
For $n$-dimensional AR(1) processes with Gaussian excitation, the compression rate is evaluated in terms of the Hausdorff dimension in \cite{lawler1996hausdorff}.

%
Continuous-domain ARMA processes in which the excitation noise is a L{\'e}vy process are studied in \cite{brockwell2005levy}.
%
%
%
%
%
%
%
The compressibility of continuous-domain innovation processes and their comparisons are provided in \cite{ghourchian2018}; the compressibility measure is based on the entropy rate of finely quantized samples.   
The study of differential entropy of finely sampled L\'evy process with continuously distributed excitation noise is achieved in \cite{fageot2020entropic}.


\subsection*{Contributions}
%
%
%
%
%

In this paper, we study the compressibility of ARMA processes, in which the excitation noise is not limited to the Gaussian or absolutely continuous probability measures, but a larger class of discrete-continuous measures. 
Accordingly, we study the information-theoretic measures of compressibility, such as block-average information dimension (BID) and the information dimension rate (IDR), originally introduced in \cite{Jalali2016uni} and \cite{koch2019}, respectively, and almost lossless compressibility measures, which is introduced in compressed sensing literature \cite{wu2010renyi}.
%
%

As we discuss in Section \ref{sec: com_measures}, BID and IDR are  similar in definition,  however, they induce two different approaches towards compressibility. On one hand, the IDR, $d_I\big(\{\Xv_t\}\big)$, of a process $\{\Xv_t\}$ quantifies the ratio of minimum number of bits needed to encode the high-resolution (i.e., $m\to\infty$) quantized version $\big\{[\Xv_t]_m\big\}$ of the process $\{\Xv_t\}$, to the minimum number of bits needed to encode the quantized version $\big\{[\Yv_t]_m\big\}$ of all encode-able processes $\{\Yv_t\}$. 
On the other hand, the BID, $d_B\big(\{\Xv_t\}\big)$ calculates the average information dimension of truncated samples of a process, for large number of samples.

Following above definitions, one can argue that the IDR encodes the definition of compressibility in a more sensible way. However, finding its value is theoretically more complex, as calculating this measure incolves finding the entropy rate of a quantized version of the stochastic process in a non-asymptotic setting for quantization step-size.

To address this issue, there are some works that show the equality of IDR and BID \textemdash where the latter is generally easier to calculate \textemdash under some conditions on the process \cite{Rezagah2016}, \cite{koch2019}. One of the conditions that generally simplify the evaluation of IDR is finiteness of mutual information among samples of the process. As we will show by an example in Section \ref{sec: BID_IDR}, this is not the case for ARMA processes with discrete-continuous excitation noise. 
Although this condition is violated for the mentioned ARMA case, as a first contribution of this work, we show that  IDR and BID are still equal.



The classical information-theoretic measures of compressibility (such as entropy and RID) search over all possible functions that can encode the data. Oftentimes, the optimal encoding function demonstrates irregular behavior in response to the input data, making the compression scheme sensitive to noise and non-idealities. this is in contrast with the compressed sensing scenario \cite{donoho2006compressed} where by restricting the encoder to linear functions, the effect of noise is kept under control.
%
Besides, under certain conditions (the encoder satisfies the restricted isometry property with a suitable constant) the standard decoder acts as a Lipschitz operator \cite{candes2005decoding} (noise and non-idealities could be linearly bounded in the output).
%
%
Drawing inspirations from the compressed sensing scenario, in this paper we mainly study the latent dimension of the optimal linear/Borel and Borel/Lipschitz encoder/decoder pairs. The minimum compression rate among all encoder/decoder pairs such that the decoding error probability is at most $\ep\in (0, 1)$ is called the minimum $\ep$-achievable compression rate (linear/Borel minimum $\ep$-achievable and Borel/Lipschitz minimum $\ep$-achievable  compression rates).
%
%
As the second contribution of this paper, we show that both minimum $\ep$-achievable rates for ARMA processes  coincide with BID and IDR (as BID and IDR are equal here). The latter equality was previously shown for an i.i.d. sequence of random variables in \cite{wu2010renyi}, and for moving-average processes in \cite{CharusaieISIT2020P1}. The result in this paper for ARMA processes extends both results.
\subsection*{Notations}
The following notations are adopted in the remainder of this work. 

\noindent
$\bullet$ \underline{\emph {Random variables (RV) and distributions:}}
the set of RVs $\{X_m,\ldots X_n\}$ is abbreviated as $\{X_i\}_{i=m}^n$.
For brevity, we define $\{X_i\} = \{X_i\}_{i=-\infty}^\infty$.  
When this set of random variables is used to construct a random vector, we employ the notation  $\Xv_m^n = [X_m,X_{m+1},\ldots, X_n]$ with $n \ge m$.  
Again, when $m=1$, the subscript is omitted, i.e. $\Xv_1^n=\Xv^n$. 
By abuse of notation, for a binary vector $\sv\in\{0, 1\}^n$, $\Xv^{\sv}$ denotes a random vector formed by the elements $X_i$ of $\Xv^n$, where $s_i=1$.
The absolute $\alpha$-moment (around zero) of the RV $X$ is denoted by $M_X(\alpha)=\Ebb[|X|^{\alpha}]$ for $\alpha\in \Rbb$. 
Further, by $M_X^0(\infty)$ we mean {\changed ${\rm ess\, sup} \, |X|$}. 
Equality in distribution is indicated as $\stackrel{\rm d}{=}$.
The discrete/continuous part of the RV $X$ is indicated as $X_{\dm}/X_{\cm}$, respectively.
The Bernoulli RV with success probability $p$ is indicated as ${\rm Bern}(p)$, the Gaussian distribution with mean $\mu$ and variance $\sgs$ are represented as  $\rm {Gauss}(\mu,\sgs)$, and the Bernoulli-Gaussian distribution is denoted by ${\rm Bern-Gauss}(p,\mu,\sgs)$.

Shannon entropy function is shown by $\Hsf(\cdot)$, the differential entropy by $\hsf(\cdot)$,  the mutual information by $\Isf(\cdot;\cdot)$ and the Kullback-Leibler divergence by $\Dsf(\cdot||\cdot)$.  For the sake of simplicity in expressing our results, we extend the notion of random vectors to $0$-dimensional random vector $\Xv$ or \emph{null}, and with an abuse of notation, we assume that in such case $\hsf(\Xv)=0$.

 The notation $\delta_x$ refers to the Dirac's measure, where {\changed 
 \ean{\delta_x(A) = \lcb \begin{array}{c c}
 1, & x\in A\\
 0, &\text{otherwise}.
 \end{array}\rnone
 }
 }

{\changed
\noindent
$\bullet$ \underline{\emph{Set theory}:}
%
Set subtraction is shown as $\Acal\setminus\Bcal=\Acal\cap\Bcal^{c}$.  The span of a set $\Acal$ of vectors is denoted as $\spanM(\Acal)$. Minkowski difference of two sets $\Acal$ and $\Bcal$ is indicated as $\Acal-\Bcal = \{\av-\bv:\av\in\Acal, \bv\in \Bcal\}$.
For an affine set $\Acal$, $\dim(\Acal)$ stands for its {\changed Euclidean} dimension.
The sets $\{i, i+1, \ldots, j\} \subseteq \Nbb$  and $\{1, \ldots, j\}$ are abbreviated as $[i:j]$ and $[j]$, respectively.	
{\changed A set $\Mcal$ of linearly dependent vectors} is called minimally dependent, if all proper subsets of $\Mcal$ are linearly independent. The set of all minimally dependent subsets of a set $\Acal$ of vectors is shown as $\Omega(\Acal)$.
}

\noindent
$\bullet$ \underline{\emph{Vectors and matrices:}}
Given an $m\times n$ matrix $A$, we denote the $i$-th column of $A$ by $A^{[i]}$, for $i\in[n]$.
In addition, for a binary vector $\sv\in\{0, 1\}^n$, $A^{[\sv]}$ denotes the sub-matrix of $A$ formed by columns $A^{[i]}$ of $A$ for $i\in [n]$, where $s_i=1$. 
The rank, spark and the span of a matrix $A$ are represented by $\rank(A)$, $\spark(A)$ and $\spanM(A)$, respectively. 
For an arbitrary matrix $A$ (not necessarily square), $\det^{+}{(A)}$ refers to the product of the non-zero singular values of  $A$.
The conjugate transpose of the matrix $A$ is indicated as $A^{\dagger}$.
The $\alpha$-vector-norm is represented by $\| \cdot \|_{\alpha}$.
For $\vv^n \in \{0,1\}^n$,  $\vov^n$ is the vector obtained by obtaining logical negation of all the elements in $\vv$.
%
%
The $n$-dimensional column vector of all zeros/ones is indicated as $\zeros_{n}$/$\ones_{n}$.
Similarly, the $n \times m$  all zeros/ones matrix is indicated as $\zeros_{n \times m}$/$\ones_{n \times m}$.

We employ the notation $I_{\text{fcr}}(A)$ as a { full column-rank} indicator function, i.e., 
\ea{I_{\text{fcr}}(A_{m\times n}) =\left\{\begin{array}{l l}
		1 & \rank(A) = n\\
		0 & \text{otherwise.}
	\end{array}\right.
}

\noindent
$\bullet$ \underline{\emph{Other notations:}}
For $\al \in [0,1]$, define $\alb=1-\al$.
%
%
\noindent
The uniform quantization of the RV $\Xv^n$ with precision $m$ is defined as
{\changed
\ea{
	[\Xv_{}^n]_m \triangleq \Big[\tfrac{ \lfloor m X_{1} \rfloor}{m}, \ldots, \tfrac{ \lfloor m X_{n} \rfloor}{m}\Big] ,
	\label{eq:uniform quantization notation}
}
}
with $[\Xv_{}^n]_m \in (\Nbb/m)^n$ and where $\lfloor x \rfloor$ is the floor of $x$.
\section{Singularity in ARMA Processes}
\label{sec:An Illustrative Example}

\begin{figure*}[t]
    \centering
    \begin{subfigure}{0.23\textwidth}
        \input{Figures/Mixtures}
        \caption{$\mu_{\xi_1}$}
        \label{fig:subfig-a}
    \end{subfigure}
    \begin{subfigure}{0.23\textwidth}
        \input{Figures/Self-Convolution}
        \caption{$\mu_{X_1}$}
        \label{fig:subfig-b}
    \end{subfigure}
    \begin{subfigure}{0.23\textwidth}
\raisebox{-3.9cm}{
\begin{tikzpicture}

\definecolor{darkgray176}{RGB}{176,176,176}

\begin{axis}[
width = \textwidth,
height = \textwidth,
tick align=outside,
tick pos=left,
x grid style={darkgray176},
xlabel={$X_1$},
xmin=-3, xmax=3,
xtick style={color=black},
y grid style={darkgray176},
ylabel={$X_2$},
ymin=-3, ymax=3,
ytick style={color=black}
]
\addplot graphics [includegraphics cmd=\pgfimage,xmin=-3, xmax=3, ymin=-3, ymax=3] {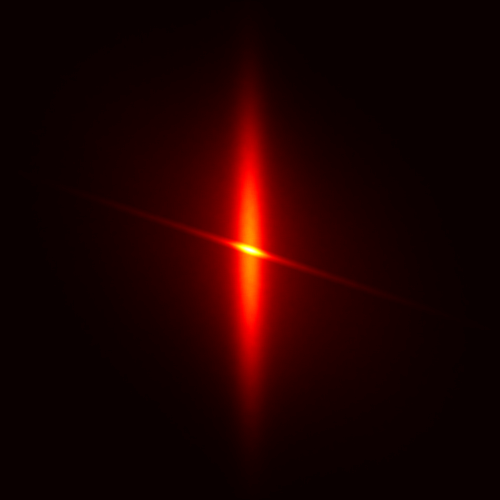};
\end{axis}

\end{tikzpicture}
}
        \caption{$\mu_{X_1X_2}$}
        \label{fig:subfig-c}
    \end{subfigure}
    \hspace{0.04\textwidth}
    \begin{subfigure}{0.17\textwidth}
    \raisebox{1cm}{
        \includegraphics[width=\linewidth, height=\linewidth]{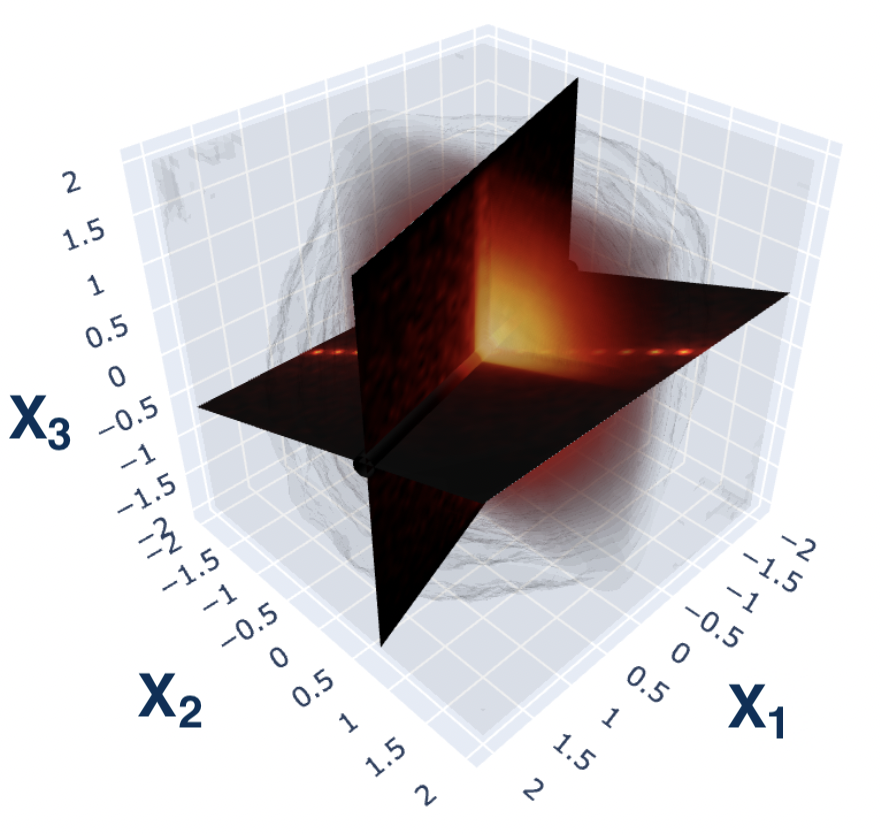}
        }
        \vspace{-.6cm}
        \caption{$\mu_{X_1X_2X_3}$}
        \label{fig:subfig-d}
    \end{subfigure}
    \caption{The distribution of $(a)$ excitation noise, $(b)$ a single sample, $(c)$ two sequential samples, and $(d)$ three sequential samples from the AR(1) process that is discussed in Section \ref{sec:An Illustrative Example}}
    \label{fig:manif}
\end{figure*}

To understand the space in which the contribution of this paper lies, we dedicate this section to illustration of the occurrence of singularities in \ac{dce} processes. Particularly, we study the marginal and joint probability measure of two examples of \ac{dce}. We show that the existence of atomic components in the excitation noise induces affine singularities \cite{charusaie22} and Cantor-type singularities in probability measure of samples.
%


These examples are concerning auto-regressive stochastic processes  that can be defined recursively as
\ea{
X_t = \xi_t +\sum_{i = 1}^{\infty} h_i X_{t-i},
\label{eq: convolutional model}
}
for which the $h_i$ quantifies the impulse response of the autoregressive filter and $\{\xi_t\}$ denotes the excitation noise. For both of the following example, we set the above filter to an AR(1) filter with $h_1 = 1/3$ and $h_i = 0$ for all other values $i \neq 1$.

{\bfseries Affine Singularity.} In the first example, assume that the excitation noise is standard Bernoulli-Gaussian (i.e., $\xi_t = b_t n_t$ where $b_t \sim Bern(1/2)$ and $n_t \sim \Ncal(0, 1)$). 
For this excitation, as we show in Lemma \ref{lem: cond_abs_cont}, the probability distribution of the sample $X_1$ is absolutely continuous. To be more precise, with a simple calculation, one can show that the distribution of $X_1$ is a uniform mixture of Gaussian distributions with variance $\sigma^2 = \frac{t}{9^k}$ for limiting case of $k\to \infty$ for $t\in T_k$ where $T_k$ is a set of either powers of $9$ or a sum of distinct powers of $9$ that are less than $9^k$.

Furthermore, due to \eqref{eq: convolutional model}, $\Xv_2^3$ can be obtained as a function of the tuple $(X_1, \xi_2, \xi_3)$ as
\eas{
X_2  & = \f {X_1} 3 +  \xi_2 
\label{eq:x2}\\
X_3 &  = \f {X_2} 3 +  \xi_3 = \f {X_1} 9+ \f {\xi_2}  3 +  \xi_3.
\label{eq:x3}
}

These identities together with the definition of $\xi_t$ illustrates the underlying distribution of $\Xv_1^3$. In fact, in case of $\xi_2 = 0$ and when $\xi_3$ takes values according the Gaussian distribution, that is an event with probability $\Pr(b_2 = 0, b_3 = 1) = 1/4$, we have $X_2 = X_1/3$. This event induces a plane on which $\Xv_1^3$ is distributed. Similarly, the event $\xi_3 = 0$ concludes in plane $X_3 = X_2 / 3$ on which the random vector is distributed. If we further assume the event $\xi_2 = \xi_3 = 0$, then we conclude that the random vector is distributed along the line $X_3 = X_2 / 3 = X_1 / 9$.

Such lower-dimensional sets are known to have zero Lebesgue measures, and we further showed that the random vector $\Xv_1^3$ lays on such sets with a non-zero probability. Such phenomenon is known as singularity in the probability measure of $\Xv_1^3$ as we further discuss in Section \ref{sec: type}. The manifestation of singularity in such AR(1) processes is demonstrated in Figure \ref{fig:manif}. As we see, while the marginal distribution of $X_1$ is absolutely continuous, the atomic singularity of $\xi_1$ leads to emergence of singularities on lower-dimensional sub-spaces in joint probability measures of $\mu_{X_1X_2}$ and $\mu_{X_1X_2X_3}$.

\begin{figure}
\centering
	\input{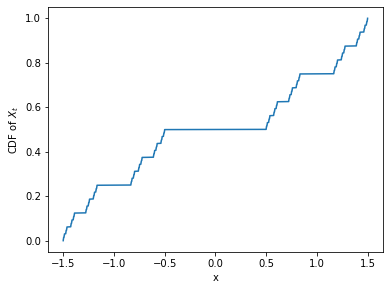}
	\caption{The CDF of a Bernoulli convolution (a realization of the AR(1) process $X_t= \xi_t+a X_{t-1}$ with Rademacher excitation noise) for $a=1/3$ which coincides with a scaled Cantor function.}\label{fig:cdf_cantor}
\end{figure}

{\bfseries Self-similar Singularity.} Another widely-studied instance of the singularity in such AR(1) process occurs when the excitation noise is a
Rademacher process \cite{kershner1935symmetric,picinbono05}, i.e.,
\ea{
\xi_k = \lcb \p{
-1  & 1/2 \\
+1  & 1/2
}\rnone.
}
For this setting $X_N$ corresponds to a  Bernoulli convolution  of order $N$, and this is shown in the limit $N\goes \infty$ to converge a scaled version of a random variable with Cantor's distribution (see Fig. \ref{fig:cdf_cantor}). 

The singularities  are well-documented in the literature but generally cover only the case of purely discrete excitation noise .  
As we shall see, the results in this paper will partially extend the study of these singularities to the case of discrete-continuous excitation noise.
%
%




\section{Preliminaries}
\label{sec:Preliminaries}

In this section, we present useful definitions of probability measures, rational functions, ARMA processes, affinely singular RVs, and information-theoretic compressibility measures.


\subsection{Types of Measures} \label{sec: type}
The cornerstone of our work are discrete-continuous probability measures. To properly introduce such measure, we start by two simpler definitions and review the well-known Lebesgue-Radon-Nikodym theorem for characterizing generic measures (see \cite[p. 121]{rudin2006real}).


 In the following, let $\Sigma$ be a $\sigma$-field of $\Rbb^n$. 
\begin{defi}[Absolutely continuous measures]
	We refer to the probability measure $\mu(\cdot)$ on $\Sigma$ as  \emph{absolutely continuous}, if for every set $S\in \Sigma$ with zero Lebesgue measure, we have that $\mu(S)=0$.
\end{defi}

\begin{defi}[Singular measures]
	A measure $\mu(\cdot)$ on $\Sigma$ is called \emph{singular}, if there exists a subset $S\subset\Rbb^n$ with zero Lebesgue measure such that
	\ea{
		\mu(\Rbb^n \setminus S)=0.
	}
	If $S$ is further countable, then, $\mu(\cdot)$ is called \emph{discrete}.
\end{defi}

\begin{Theorem}[Lebesgue-Radon-Nikodym] \label{thm: lrn}
	Every probability measure $\mu$ on $\Rbb^n$ is associated with a unique singular measure $\mu_s(\cdot)$ and an absolutely continuous measure $\mu_a(\cdot)$, such that for a $0\leq p\leq 1$ we have
	\ea{ \mu =p\,\mu_a+(1-p)\,\mu_s.
	}
We refer to $p$ as \emph{continuity chance} of the measure $\mu(\cdot)$.
	%
If 	$\mu_s(\cdot)$ is discrete and $0 < p < 1$ (no equality), then, we call $\mu$ a $p$-discrete-continuous probability measure.
\end{Theorem}


\subsection{Rational functions and Hankel matrices}
\begin{defi}
	The function $R:\Cbb\,\to\,\Cbb$ is rational, if there exists $h,g\in\mathbb{C}[x]$ (the set of all polynomials with complex-valued coefficients) with $g\neq 0$ such that
	\ea{
		R(z) = \frac{h(z)}{g(z)}.
	}
%
The function is further called \lq\lq proper\rq\rq ~ if ${\rm deg}(g)\geq {\rm deg}(h)$. 
	The roots of $g$ and $h$ are commonly referred to as the poles and zeros of $R$ with their multiplicities. If $p$ is simultaneously a root of both $g$and  $h$ with multiplicities $k_g$ and $k_h$ where $k_h\geq k_g$, then $p$ is called a removable pole.
%
\end{defi}

%
%

\begin{defi} [Zeros and Poles]	 
	We say $+\infty$ (or $-\infty$) is a zero of $R(\cdot)$ if $\lim_{z\to+\infty} g(z)=0$  (or $\lim_{z\to-\infty} h(z)=0$). A similar statement holds for the poles at infinity.
\end{defi}

\begin{rem}\label{rem:one}
	A rational function is proper, if and only if it has no $\pm \infty$ poles. 
\end{rem}

\begin{rem}\label{rem:closed}
	The set of rational functions are closed under  multiplication and linear combination.
\end{rem}

\begin{lemma}[\cite{gantmakher1959theory} Chapter V, Theorem 8 \& Corollary in p.245]\label{lem: hankel}
	Let $H(z)$ be a proper rational function with $p$ non-removable poles (including multiplicities). 
	If $h[n]$ is such that $\sum_{n=0}^{\infty} h[n]z^{-n} = H(z)$ (i.e., $h[n]$s are the Laurent series of $H(z)$  around $z=0$, or the causal inverse $z$-transform of $H(z)$), then, the Hankel matrix $A=\big[h[i+j]\big]_{i, j=0}^{p-1}$ has non-zero determinant.
%
\end{lemma}

\subsection{Compressibility Measures}\label{sec: com_measures}

We now briefly review some of the compressibility measures of random processes and sequences of random variables.

The classical notion of entropy is well-defined for discrete-valued RVs. For continuous-valued RVs, this notion is generalized via the limiting entropy of the quantized RV. 
The uniform quantization of a RV $\Xv^n$ with precision $m$ is defined in \eqref{eq:uniform quantization notation} as $[\Xv^n]_m\in (\Nbb/m)^n$.

\begin{defi}[{\cite{renyi1959dimension}}]
	\label{def:RID}
	The \emph{R\'enyi information dimension} (RID) for a RV $\Xv^n$  is 
	\ea{
		d(\Xv^n)=\lim_{m\to\infty}\tfrac{H([\Xv^n]_m)}{\log m},
	}
	if the limit exists, where $H(\cdot)$ is the Shannon entropy function. 
\end{defi}

\begin{Theorem}[\cite{renyi1959dimension}]\label{thm:disc_cont_dim}
	For a random variable $X$ with $p$-discrete-continuous measure, we have $d(X)=p$. 
\end{Theorem}

\begin{defi}[\cite{Jalali2016uni}]
	For a discrete-domain  stationary stochastic process $\{\Xv_t\}$, the block-average information dimension (BID) is defined as
	\ea{
		d_B\big(\{\Xv_t\}\big)=\lim_{n\to\infty}\lim_{m\to\infty} \f {H\big([X_n]_m|[\Xv_{}^{n-1}]_m\big)}{\log m}	,
		\label{def:BID}
	}
	if the limit exists. If $\lim_{m\to \infty}$ in \eqref{def:BID} does not exist,  $\overline{d}_B\big(\{\Xv_t\}\big)$ and $\underline{d}_B\big(\{\Xv_t\}\big)$ represent \eqref{def:BID} when $\lim_{m\to \infty}$ is replaced with  $\limsup_{m\to\infty}$ and $\liminf_{m\to \infty}$, respectively.
\end{defi}

\begin{Theorem}{\cite[Lem. 3]{Jalali2016uni}}
	\label{thm:BID sss}
	For a discrete-domain stationary stochastic process $\{\Xv_t\}$ the BID equals
	\ea{
		d_B\big(\{\Xv_t\}\big)
		& =\lim_{n\to\infty}\lim_{m\to\infty}\f {H\big([\Xv^n]_m\big)}{n\log m} \nonumber \\
		& = \lim_{n\to\infty}\f 1 n d(\Xv^n).
\label{eqn: d_B_limit}
	}
\end{Theorem}

\begin{defi}[\cite{koch2019}]
	\label{def:IDR}
	For a discrete-time stochastic process $\{\Xv_t\}$, \emph{information dimension rate} (IDR) is defined as
	\ea{
		d_I\big(\{\Xv_t\}\big)=\lim_{m\to\infty}\lim_{n\to\infty}\f {H\big([\Xv_{}^n]_m\big)}{n\log m},
		\label{eq:IDR}
	}
	if  $\lim_{m\to\infty}$ exists; otherwise, by replacing $\lim_{m\to\infty}$ with $\liminf_{m\to\infty}$ and $\limsup_{m\to\infty}$, we can define $\underline{d}_I\big(\{\Xv_t\}\big)$ and $\overline{d}_I\big(\{\Xv_t\}\big)$, respectively.
\end{defi}

\begin{Theorem}[Thm. 14,\cite{koch2019}]
	\label{th:koch2019}
	For a discrete-domain stationary stochastic process $\{\Xv_t\}$, we have that
	\ea{
		d_I\big(\{\Xv_t\}\big)\leq d_B\big(\{\Xv_t\}\big).\label{eqn: d_B_d_I}
	}
\end{Theorem}

\begin{defi}[{\cite{koch2019}}]
	\label{def:RDF}
	The \emph{quadratic rate-distortion function} (QRDF) of a RV $\Xv^n$ is defined as
	\ea{
		R_2(\Xv_{}^n, D)=\inf_{\mu_{\Xhv_{}^n|\Xv_{}^n}(\cdot, \cdot): \ \Ebb\|\Xv^n-\Xhv^n\|_2^2 <D} I(\Xv_{}^n; \Xhv_{}^n),
		\label{eq:QRDF}
	}
	where $\mu_{\Xhv_{}^n|\Xv_{}^n}$ is the conditional probability measure of $\Xhv_{}^n$ given $\Xv_{}^n$.
\end{defi}

%

\begin{defi}[{\cite{Dembo1994}}]
	\label{def:RDD}
	For a RV $\Xv^n$, the \emph{rate-distortion dimension} (RDD) is defined as
	\ea{
		d_{R}(\Xv_{}^n)=2\lim_{D\to 0^{+}}\f {R_2(\Xv_{}^n, D)}{-\log D},
	}
	if the limit exists.
\end{defi}

The smooth and robust compression/decompression methods are studied in \cite{smoothcomp} and \cite{wu2010renyi}. For the following compressibility measures, we use the notations introduced in \cite{wu2010renyi} which are inspired by similar notions in the field of compressed sensing \cite{https://doi.org/10.1002/cpa.20124}.

\begin{defi}\label{def: epsilon-rate}[\cite{wu2010renyi}] Let $\{\Xv_t\}$ be a discrete-domain stochastic process. For an integer $n$, we refer to $f_n:\Rbb^n\to\Rbb^{\lfloor nR_n\rfloor}$ and $g_n:\Rbb^{\lfloor nR_n\rfloor}\to \Rbb^{n}$ as an $\ep$-encode/decode pair with rate $R_n$ (for the process), once we have
	\ea{
	\Pr\Big(g_n\big(f_n(\Xv^n)\big)\neq \Xv^n\Big)\leq \ep.	
	}
For a given $\ep >0$, 
we define the minimum $\ep$-achievable rate as $\liminf_{n\to\infty} R_n$. Further, if we restrict the encoder $f_n$ to be linear, we obtain the linear-encode $\ep$-achievable rate  represented by $R^{*}(\ep)$.
Alternatively, if we restrict the decoder to be Lipschitz, then, we obtain the minimum Lipschitz-decode $\ep$-achievable rate  denoted by $R(\ep)$. 
	
\end{defi}




\subsection{Stochastic Processes}
\label{sec:Stochastic Processes}
A stationary process is a  stochastic process whose unconditional joint probability distribution does not change when shifted in time.

\begin{defi}[Stationary process]
	A stochastic process $\{\Xv_t\}$ is  \lq\lq strictly stationary\rq\rq~ if 
	$$\mu_{X_{i_1+l}, \ldots, X_{i_n+l}}(\cdot)= \mu_{X_{i_1}, \ldots, X_{i_n}}(\cdot),$$
	for every $n\in \Nbb$ and $l\in \Zbb$.
\end{defi}

The class of \ac{arma} processes is defined as following:

\begin{defi}[\ac{arma} Process \cite{box2015time}]
\label{def:ARMA Process}
	A $(p,q)$-\ac{arma} process with $ p,q \in \Nbb$  is  defined by the recursive expression
	\ea{
		X_t = \xi_t +\sum_{i \in [q]} \theta_i \xi_{t-i} -\sum_{i \in [p]} \phi_{i} X_{t-i},	
		\label{eqn: arma_eqn}
	}
 for $t \in \Zbb$ and 	where constants $\{\phi_i \}_{i\in[p]}$/$\{\theta_i \}_{i\in[q]}$  are  moving-average (MA) and autoregressive (AR) parameters, respectively.  
 %
	The sequence $\{\xi_i\}_{i \in \Zbb }$ is defined as the excitation process and is a sequence of i.i.d. RVs.

	Equivalently, for any $n \geq p+1$, the process in \eqref{eqn: arma_eqn} can be expressed through the vector equality 
	\ea{
		%
		\Phi\cdot\Xv^{n} = 
		\The \cdot
		\xiv^{n}_{p-q+1},
		\label{eq:vector expression}
	}
 where $\Phi \in \Rbb^{(n-p) \times n}$ and $\The \in \Rbb^{n-p \times (n+q-p)}$ are  Toeplitz matrices defined as 
	\ea{
		\Phi &= \left[\begin{array}{c c c c c c c}
			\phi_p & \phi_{p-1} &\ldots& \phi_1& 1 &\ldots & 0\\
			\ldots & \ddots & \ddots & \ddots & \ddots & \ddots & \ldots\\
			\ldots & 0 &  \phi_p & \phi_{p-1} &\ldots& \phi_1& 1
		\end{array}\right],
		\label{eq:phi}
	}
	and
	\ea{
		\The &= \left[\begin{array}{c c c c c c c}
			\theta_q & \theta_{q-1} &\ldots& \theta_1& 1 &\ldots & 0\\
			\ldots & \ddots & \ddots & \ddots & \ddots & \ddots & \ldots\\
			\ldots & 0 &  \theta_q & \theta_{q-1} &\ldots& \theta_1& 1
		\end{array}\right].
		\label{eq:theta}
	}
 \end{defi}

Note that, in Definition \ref{eqn: arma_eqn}, the process is defined over $\Zbb$. 
Accordingly, a set of samples from the process for $t \in [n,m]$ is expressed in vector forms as $\Xv_n^m$ for $n<m$, $n,m \in \Zbb$.

%
%
%
%
%
%

An alternative representation of the \ac{arma} process in Definition \ref{def:ARMA Process} through the  $\mathcal{Z}$-transform as the
%
filtered version of the excitation process, 
%
through the filter 
	\ea{
		H(z)=\displaystyle \f{1+ \sum_{i \in  [q]} \theta_i z^{-i}}{1 +\sum_{i \in [p]} \phi_i z^{-i}},
		\label{eq:filter}
	}
	where $H(z)$ stands for the $\mathcal{Z}$-transform of the filter's impulse response. 
 

	One can also rewrite this rational function in the canonical form as
	\ea{
		H(z)=\frac{\Pi_{i=1}^{n_n} (r_iz^{-1}-1)^{z_i}}{\Pi_{i=1}^{n_d}(a_iz^{-1}-1)^{p_i}}	,\label{eqn: transfer}
	}
	where the pairs $a_i$/$p_i$ describe the filter  pole/zero  while $r_i$/$z_i$ its multiplicity. 
	%
	%
 
	\begin{rem}
	A stationary \ac{arma} process is  an \ac{arma}$(p, q)$ process that is strictly stationary. It is shown in \cite{armastationary} that an \ac{arma}$(p, q)$ process with the excitation noise $\{\xi_i\}$ is stationary if and only if one of the below conditions hold:
 
	\begin{enumerate}
	\item $H(z)$ is well-defined for all $z\in\mathbb{C}\setminus \{0\}$,
	
	\item $H(z)$ is bounded for all $|z|=1$ and \ea{\Ebb\bigg[\max\Big\{0, \log \big(|\xi_1|\big)\Big\}\bigg]<\infty.
 }
	
	\end{enumerate}
%
%
%
	\end{rem}


When the excitation white noise in an \ac{arma} process has a discrete-continuous distribution, the vectors of \ac{arma} samples might also have singular components in their distribution. 
Due to the linear mixture of the white noise samples, these components form affine subsets.
The class of vectors with with singularities over affine sets was the topic of our previous publication \cite{charusaie22}. 
We refer to this class of vectors as Affinely Singular Random Vectors (\ac{asrv}). 




\begin{defi}[\cite{CharusaieISIT2020P1}]\label{def:aff_sing}
A random vector $\Zv^n$ is an \ac{asrv} if there exists a finite 
or countably infinite
number of affine subsets $\Acal_i \subset \mathbb{R}^n$ of dimension $0\leq d_i \leq n$ and absolutely continuous \footnote{Here, we abused the notion of absolute continuity. In fact, if $f_i$ is the affine transformation with which we can generate the set $\Acal_i$, then, by the absolute continuity over $\Acal_i$ we mean the absolute continuity with respect to the pushforward measure $\ell_i(f^{-1}_i(\cdot))$, where $\ell_i$ is a $d_i$-dimensional Lebesgue measure.} $d_i$-dimensional measures $\mu_i$ over $\Acal_i$ such that 
\begin{align}
\forall\, \Bcal\subseteq \mathbb{R}^n:~ \mathbb{P}\big(\Zv^n\in \Bcal\big) = \sum_{i} \mu_i\big( \Bcal \cap \Acal_i\big).
\end{align}
\end{defi}

We finally come to a definition of \ac{dce} processes; a stationary ARMA process in which the excitation noise is discrete-continuous, so that samples of the process are \ac{asrv}.

\begin{defi}\label{def: DC_ARMA}
A \ac{dce} process is defined as the  $(p, q)$-ARMA  processes  in which the excitation process $\{\xi_t\}$ has 
a discrete-continuous distribution.
Such a process is indicated as 
$(p,q)$-\ac{dce} process. 
\end{defi}

In the next section we shall focus on the study of the compressibility of \ac{dce} processes as measure in terms of the compressibility measures of Sec. \ref{sec: com_measures}.

\section{Main Results}
\label{sec:Main Contribution}
Having formally introduced the \ac{dce} processes in Sec. \ref{sec:Stochastic Processes}, we focus on studying the structure of their singularities through the compressibility measures in Section \ref{sec: BID_IDR} and \ref{sec: ep_achievable}.
The main result of these sections is described, from a high-level perspective, as follows:

 $\bullet$  If the excitation noise is purely discrete, then 
%
the  distribution of truncated samples of the resulting \ac{dce} process contains a singularity such that its compressibility measure of  Section \ref{sec: BID_IDR} and \ref{sec: ep_achievable} behave similar to that of a purely discrete measure. This means that asymptotically the necessary information for reconstructing the discretized version of the process is concentrated within a diminishing number of bits per sample, as we increase the number of samples. At the same time, the more samples of the process we want to linearly compress and robustly recover, the less the ratio of the hidden variables that we need. 
%
%

$\bullet$ When the excitation process process is purely continuous, then the truncated \ac{dce} process  has a continuous distribution. This means that theoretically it is impossible to compress this process with a non-trivial ratio.

$\bullet$ We prove that when the excitation process is discrete-continuous  
then, the truncated \ac{dce} process  has  affinely singular distribution. Further, we show that the average dimension of such singularities concentrates around the continuity chance of the excitation noise. This helps us to prove that the compressibility of the resulting process is equal to that of the excitation noise too.

To summarize, the main message of the following theoretical results can be summarized as following: 

`` \emph{An \ac{dce} process preserves the information theoretical compressibility and smooth and robust compressibility of its excitation noise, both equal to the discrete probability of the excitation law.} "

We start our analysis by obtaining information-theoretical measure of compressibility for a \ac{dce} process in the following section.

\subsection{BID and IDR of a DCE-ARMA process}\label{sec: BID_IDR}

To study the block-average information dimension of a \ac{dce} process, we need to use the calculus of information dimensions. This calculus that is summarized in \cite{WuThesis} embodies the following properties that we use in our proofs:

\vspace{0.2cm}
 
    1) The information dimension of a random vector $\Xv^m$ is bounded above by $m$.
    
    2) A Lipschitz function increases the information dimension of a variable.
    
    3) For two independent random vectors $\Xv^{m}$ and $\Yv^{n}$, the joint information dimension is equal to the sum of information dimension of each, i.e., $d(\Xv^{m}, \Yv^{n})=d(\Xv^m)+d(\Yv^n)$.
    
    4) For two independent random vectors $\Xv^m$ and $\Yv^m$, the information dimension of the sum of random vectors is bounded below by the information dimension of each, i.e., $d(\Xv^m+\Yv^m)\geq \max\{d(\Xv^{m}), d(\Yv^m)\}$.

\vspace{0.2cm}

 Using these rules, we prove in the next theorem that the BID of all stationary ARMA processes, whether the excitation noise is discrete-continuous or not, are equal to the information dimension of their excitation noise.
%
Under some mild conditions, we further show that 
 the IDR, the BID and the RID of excitation noise samples coincide. 
 
\begin{Theorem}\label{thm: ARMA_inf_dim}
If $d(\xi_1)$ is well-defined for the i.i.d. excitation process $\{\xi_t\}$, then, for the resulting stationary ARMA process $\{\Xv_t\}$ we have
%
	\ea{
		d_B\big(\{\Xv_t\}\big)=d(\xi_1).
}
 Further, if $H\big([\xi_1]_1\big)<\infty$, then,
\ea{
d_R\big(\{\Xv_t\}\big)=d_I\big(\{\Xv_t\}\big)=d_B\big(\{\Xv_t\}\big)=d(\xi_1).
}
\end{Theorem}

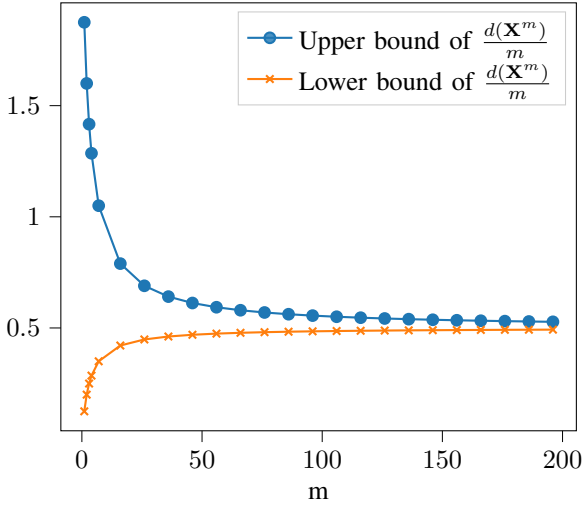
\begin{figure}
	\centering
\begin{tikzpicture}

\definecolor{darkorange25512714}{RGB}{255,127,14}
\definecolor{lightgray204}{RGB}{204,204,204}
\definecolor{steelblue31119180}{RGB}{31,119,180}

\begin{axis}[
legend cell align={left},
legend style={fill opacity=0.8, draw opacity=1, text opacity=1, draw=lightgray204},
tick align=outside,
tick pos=left,
x grid style={white},
xlabel={m},
xmajorgrids,
xmin=-8.75, xmax=205.75,
xtick style={color=black},
y grid style={white},
ymajorgrids,
ymin=0.0375, ymax=1.9625,
ytick style={color=black}
]
\addplot [thick, steelblue31119180, mark=*, mark size=2, mark options={solid}]
table {%
1 1.875
2 1.6
3 1.41666666666667
4 1.28571428571429
7 1.05
16 0.789473684210526
26 0.689655172413793
36 0.641025641025641
46 0.612244897959184
56 0.593220338983051
66 0.579710144927536
76 0.569620253164557
86 0.561797752808989
96 0.555555555555556
106 0.55045871559633
116 0.546218487394958
126 0.542635658914729
136 0.539568345323741
146 0.536912751677852
156 0.534591194968553
166 0.532544378698225
176 0.53072625698324
186 0.529100529100529
196 0.527638190954774
};
\addlegendentry{Upper bound of $\frac{d(\mathbf{X}^m)}{m}$}
\addplot [thick, darkorange25512714, mark=x, mark size=2, mark options={solid}]
table {%
1 0.125
2 0.2
3 0.25
4 0.285714285714286
7 0.35
16 0.421052631578947
26 0.448275862068966
36 0.461538461538462
46 0.469387755102041
56 0.474576271186441
66 0.478260869565217
76 0.481012658227848
86 0.48314606741573
96 0.484848484848485
106 0.486238532110092
116 0.487394957983193
126 0.488372093023256
136 0.489208633093525
146 0.48993288590604
156 0.490566037735849
166 0.49112426035503
176 0.491620111731844
186 0.492063492063492
196 0.492462311557789
};
\addlegendentry{Lower bound of $\frac{d(\mathbf{X}^m)}{m}$}
\end{axis}

\end{tikzpicture}
 \caption{Lower-bound and upper-bound of the average RID of an AR(2) process with $d(\xi_1)=0.5$ that is derived in Theorem \ref{thm: ARMA_inf_dim}, and in terms of the number of samples of the process.}\label{fig: bid_arma}
\end{figure}

Although we postpone the full proof of this theorem to Appendix \ref{app: ARMA_inf_dim}, we provide the reader with the sketch of the part related to the BID in the following to familiarize the reader with the techniques employed. Firstly, due to the recurrence relation of the ARMA process $\{\Xv_t\}$, we prove that a truncation  $\Xv^{m+p}$ of this process, where $m$ is an arbitrary integer, and $p$ is the number of poles of the process, can be linearly obtained from the excitation noise $\xiv_{1-q+p}^{m+p}$ and a smaller truncation $\Xv_{1}^{p}$. As a result of Statement 2 in above, the information dimension can only increase $d(\Xv^{m+p})\leq d(\xiv_{1-q+p}^{m+p}, \Xv_{1}^{p})$, where then using the first and the third rule, can again be upper-bounded by $d(\xiv_{1-q+p}^{m+p})+ d(\Xv_{1}^{p})\leq (m+q)d(\xi_1) + p$. Furthermore, once more due to the linear dependence of process and the excitation noise, and as a result of rule 4, we can lower-bound the information dimension as $d(\Xv^{m+p})\geq d(\xiv_{1+p}^{m+p}) = md(\xi_1)$. As a result, the limit $\lim_{m\to\infty} \frac{d(\Xv^{m+p})}{m+p}$ that quantifies the BID is equal to $d(\xi_1)$. The upper- and lower-bound of this proof is illustrated for an AR(2) process in Figure \ref{fig: bid_arma}.

\begin{figure}
\begin{minipage}[b]{0.48\linewidth}	
\includegraphics[width=0.9\linewidth]{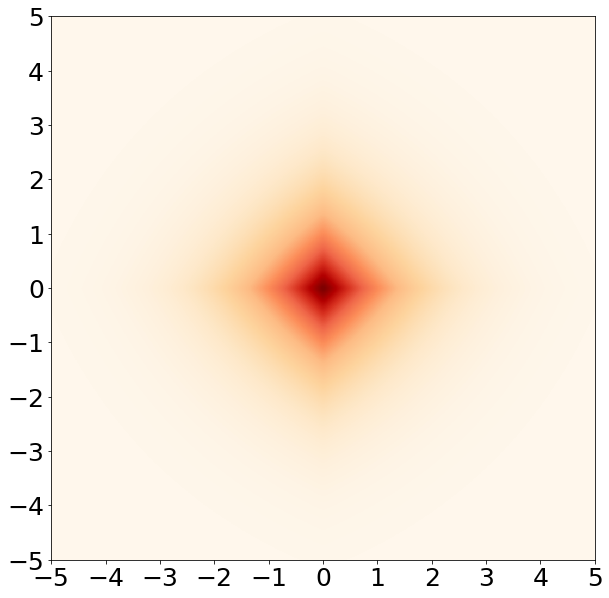}
\centerline{(a)}\medskip
\end{minipage}
\begin{minipage}[b]{0.48\linewidth}	
\includegraphics[width=0.95\linewidth]{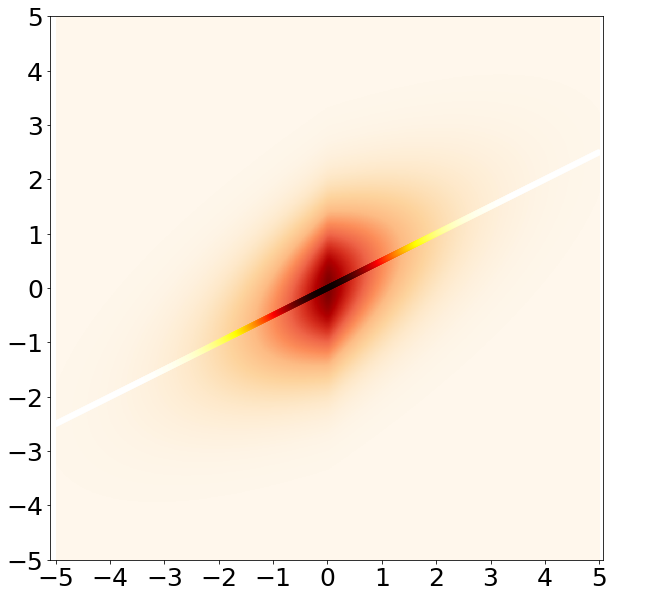}
\centerline{(b)}\medskip
\end{minipage}
  \caption{(a) Product measure of two samples of an AR(1) process generated by Bernoulli Gaussian excitation noise with the RID 0.5 and $\phi_1=0.5$, and (b)  their joint measure.}
	\label{fig:AR1_disc_cont}
 \end{figure}

Here, we should mention that the technique that is used in proving the equality of $d_R\big(\{X_t\}\big)=d_B\big(\{X_t\}\big)$ and its consequences varies from the ones investigated in \cite{koch2019} and \cite{Rezagah2016}. These works obtained sufficient conditions on mutual information among samples of the process that concludes this equality, while these conditions do not necessarily hold in our setting. In particular, \cite{koch2019} shows that the equality holds when 
%
	there exists a nonnegative integer $n$ for which
	\ea{
		I(\Xv_1^k;\Xv_{-\infty}^{-n})<\infty, ~~k=1, 2, \infty.
	}
	However, this condition is not necessarily satisfied in the case of \ac{dce} processes. To see this, let  $\{\Xv_t\}$ be an $AR(1)$ process; using the Markovity of the process, we have that
	\ea{
		I(\Xv_1^k;\Xv_{-\infty}^{-n})=I(X_1; X_{-n}).
	}
	Moreover, we know 
	\ea{
		X_1  
&= {  \sum_{k=-n+1}^{1}(-\phi_1)^{1-k}\xi_k+(-\phi_1)^{n+1}X_{-n} } \nonumber \\
& = U + (-\phi_1)^{n+1}X_{-n},
	}
	where $U$ is a discrete-continuous RV with $d(U)=1-(1-d)^n$ using \cite[Lem. 11]{ghourchian2018}. 
Now, if we let $U$ have a mass probability at 	$x_d$ equal to $\Pr(U=x_d)=p = (1-d)^n>0$,
 the joint probability measure $\mu_{X_1X_{-n}}(\cdot)$ will have a nonzero probability $p$ on the line $X_1-(-\phi_1)^{n+1}X_{-n}=x_d$. Besides, using Lemma \ref{lem: cond_abs_cont} we know that $X_1$ and $X_{-n}$ both have absolutely continuous measures. Hence, the product measure $\mu_{X_1}\times\mu_{X_{-n}}(\cdot)$ is  absolutely continuous. As a result, $\mu_{X_1X_{-n}}(\cdot)$ has singularity w.r.t the measure $\mu_{X_1}\times\mu_{X_{-n}}(\cdot)$ which prevents the Radon-Nikodym derivative and the mutual information from being well-defined (see Figure \ref{fig:AR1_disc_cont} for illustration). 
 As a result of the above discussion, Theorem \ref{thm: ARMA_inf_dim} cannot be posed as an instance of the the results of \cite{koch2019} and \cite{Rezagah2016}. 

\subsection{Smooth and robust compressibility of a DCE-ARMA process} \label{sec: ep_achievable}
In previous section, we showed that the extent to which an ARMA process can be compressed is directly connected to the RID of each instance of its excitation noise. 
This compressibility rate is determined by the Information Dimension Rate (IDR), an extension of Shannon's entropy. Similar to entropy, IDR evaluates all potential compression functions for the process.
%
As an alternative to that approach, the compressed sensing field focuses on such compression and decompression pairs of functions and assumes linearity in former and smoothness in latter to make the system simple and robust to noise.

In this section, we extend our results to the cases of linear compression and smooth decompression, and quantify minimum $\ep$-achievable rate $R(\ep)$ as in Definition \ref{def: epsilon-rate} for \ac{dce} processes. To achieve that, we first need to generalize Theorem \ref{thm: ARMA_inf_dim} to further specify the probability distribution of each truncation of an ARMA process. In particular, Theorem \ref{thm: ARMA_inf_dim} showed that the  BID of the process is $d(\xi_1)$. In case that each truncation of the process has an affinely singular probability distribution, then as shown in \cite[Lemma 2]{charusaie22} such result is equivalent to the average dimension of the singular components of probability measure converge to $d(\xi_1)$. In the following theorem, we show that the dimensions of such components further does not only have the average equal to $d(\xi_1)$ but further is concentrated around that value.



\begin{Theorem}\label{eqn: arma_is_affine}
	For a stationary $(p,q)$-\ac{dce} process in 
Definition \ref{def: DC_ARMA},  if the excitation noise $\{\xi_i\}$ is $\alpha$-discrete-continuous, then, the vector of truncated samples $\Xv^n$  is  affinely singular  when  $n\geq p+1$. 
  
  In addition, the affinely singular distribution of $\Xv^{n}$ can be expressed as 
		\ea{
			\Xv^{n} \overset{d}{=} S_V [\Yv_i; 0_{n-\tilde{d}_V}]+\ev_V,	\label{eqn: aff_arma}
		}
		where $S_i$s are unitary $n\times n$ matrices, $V$ is a discrete RV on $\Nbb$ defined in \cite[Lemma 6 for $A=I$]{CharusaieISIT2020P1}, and $\ev_i$s are fixed $n$-dimensional vectors. Furthermore, $\Yv_i$s are $d_i$-dimensional absolutely continuous RV and 
		\ea{
			\Pr (d_i> k)\geq 1-\exp\Big(-(n+q-p)D\big(\tfrac{k+q-p}{n+q-p}\|\alpha\big)\Big), \label{eqn: conc_more}
		}
		for $\tfrac{k+q-p}{n+q-p}<\alpha$ and 
		\ea{
			\Pr (d_i< k)\geq 1-\exp\Big(-(n+q-p)D\big(\tfrac{k-p}{n-p}\|\alpha\big)\Big),\label{eqn: conc_less}
		}
		for $\tfrac{k-p}{n-p}>\alpha$.
		
	
\end{Theorem}

\begin{IEEEproof}
	See Appendix \ref{app: arma_affine}.
\end{IEEEproof}


\begin{figure}
	\centering
\begin{tikzpicture}{scale = 0.8}

\definecolor{brown}{RGB}{165,42,42}
\definecolor{darkgray176}{RGB}{176,176,176}
\definecolor{lightgray204}{RGB}{204,204,204}
\definecolor{paleturquoise}{RGB}{175,238,238}
\definecolor{violet}{RGB}{238,130,238}
\pgfplotsset{
  /pgfplots/rect legend/.style 2 args={
    legend image code/.code={
      \draw[##1,no markers] (0cm,-.1cm) rectangle (0.5cm,0.1cm)
        node[pos=0.5, #1] {}
        node[pos=0.5, #2] {};
    }
  }
}

\begin{axis}[
legend cell align={left},
legend style={
  fill opacity=0.8,
  draw opacity=1,
  text opacity=1,
  at={(0.03,0.97)},
  anchor=north west,
  draw=lightgray204
},
tick align=outside,
tick pos=left,
x grid style={darkgray176},
xlabel={Normalized Singularity Dimensions},
xmin=-0.039, xmax=1.039,
xtick style={color=black},
y grid style={darkgray176},
ylabel={Density of Dimensions},
ymin=0, ymax=5.67753689897313,
ytick style={color=black}
]
\draw[draw=black,fill=paleturquoise,opacity=0.3] (axis cs:0.340327017994275,0) rectangle (axis cs:0.9085381267712,1.75990927412965);
\addlegendimage{rect legend={above}{below},draw=black,fill=paleturquoise,opacity=0.3}
\addlegendentry{CI, $n=12$}

\draw[draw=black,fill=violet,opacity=0.3] (axis cs:0.427076011175366,0) rectangle (axis cs:0.795005216492022,2.71791416813285);
\addlegendimage{rect legend={above}{below},draw=black,fill=violet,opacity=0.3}
\addlegendentry{CI, $n=32$}

\draw[draw=black,fill=brown,opacity=0.3] (axis cs:0.472385336285866,0) rectangle (axis cs:0.739861681709806,3.73864835940926);
\addlegendimage{rect legend={above}{below},draw=black,fill=brown,opacity=0.3}
\addlegendentry{CI, $n=62$}

\draw[draw=none,fill=paleturquoise] (axis cs:0.00999999999999999,0) rectangle (axis cs:0.0366666666666667,0);
\addlegendimage{rect legend={above}{below},draw=none,fill=paleturquoise}
\addlegendentry{n=12}

\draw[draw=none,fill=paleturquoise] (axis cs:0.11,0) rectangle (axis cs:0.136666666666667,0);
\draw[draw=none,fill=paleturquoise] (axis cs:0.21,0) rectangle (axis cs:0.236666666666667,0.00329089829952167);
\draw[draw=none,fill=paleturquoise] (axis cs:0.31,0) rectangle (axis cs:0.336666666666667,0.0271499109710538);
\draw[draw=none,fill=paleturquoise] (axis cs:0.41,0) rectangle (axis cs:0.436666666666667,0.135749554855269);
\draw[draw=none,fill=paleturquoise] (axis cs:0.51,0) rectangle (axis cs:0.536666666666667,1.55772614196421);
\draw[draw=none,fill=paleturquoise] (axis cs:0.61,0) rectangle (axis cs:0.636666666666667,1.92424994007344);
\draw[draw=none,fill=paleturquoise] (axis cs:0.71,0) rectangle (axis cs:0.736666666666667,2.47403563723728);
\draw[draw=none,fill=paleturquoise] (axis cs:0.81,0) rectangle (axis cs:0.836666666666667,2.31940840990995);
\draw[draw=none,fill=paleturquoise] (axis cs:0.91,0) rectangle (axis cs:0.936666666666667,1.55838950668929);
\draw[draw=none,fill=violet] (axis cs:0.0366666666666667,0) rectangle (axis cs:0.0633333333333333,8.85460824262447e-11);
\addlegendimage{rect legend={above}{below},draw=none,fill=violet}
\addlegendentry{n=32}

\draw[draw=none,fill=violet] (axis cs:0.136666666666667,0) rectangle (axis cs:0.163333333333333,3.68763995589474e-07);
\draw[draw=none,fill=violet] (axis cs:0.236666666666667,0) rectangle (axis cs:0.263333333333333,0.000127949268965153);
\draw[draw=none,fill=violet] (axis cs:0.336666666666667,0) rectangle (axis cs:0.363333333333333,0.00934892313536575);
\draw[draw=none,fill=violet] (axis cs:0.436666666666667,0) rectangle (axis cs:0.463333333333333,0.199342643069134);
\draw[draw=none,fill=violet] (axis cs:0.536666666666667,0) rectangle (axis cs:0.563333333333333,2.46739412419151);
\draw[draw=none,fill=violet] (axis cs:0.636666666666667,0) rectangle (axis cs:0.663333333333333,4.09086763665561);
\draw[draw=none,fill=violet] (axis cs:0.736666666666667,0) rectangle (axis cs:0.763333333333333,2.65825148135252);
\draw[draw=none,fill=violet] (axis cs:0.836666666666667,0) rectangle (axis cs:0.863333333333333,0.546738206097449);
\draw[draw=none,fill=violet] (axis cs:0.936666666666667,0) rectangle (axis cs:0.963333333333333,0.0279286673769142);
\draw[draw=none,fill=brown] (axis cs:0.0633333333333333,0) rectangle (axis cs:0.09,6.28688806645066e-18);
\addlegendimage{rect legend={above}{below},draw=none,fill=brown}
\addlegendentry{n=62}

\draw[draw=none,fill=brown] (axis cs:0.163333333333333,0) rectangle (axis cs:0.19,1.50485075826079e-11);
\draw[draw=none,fill=brown] (axis cs:0.263333333333333,0) rectangle (axis cs:0.29,4.90686985952763e-07);
\draw[draw=none,fill=brown] (axis cs:0.363333333333333,0) rectangle (axis cs:0.39,0.000834008921574049);
\draw[draw=none,fill=brown] (axis cs:0.463333333333333,0) rectangle (axis cs:0.49,0.126873837069594);
\draw[draw=none,fill=brown] (axis cs:0.563333333333333,0) rectangle (axis cs:0.59,3.1458481432568);
\draw[draw=none,fill=brown] (axis cs:0.663333333333333,0) rectangle (axis cs:0.69,5.40717799902203);
\draw[draw=none,fill=brown] (axis cs:0.763333333333333,0) rectangle (axis cs:0.79,1.28981509054716);
\draw[draw=none,fill=brown] (axis cs:0.863333333333333,0) rectangle (axis cs:0.89,0.0294210336618712);
\draw[draw=none,fill=brown] (axis cs:0.963333333333333,0) rectangle (axis cs:0.99,2.93968189345407e-05);
\end{axis}

\end{tikzpicture}
 \caption{The concentration of normalized singularity dimensions for a (2, 3)-DCE-ARMA process with $d(\xi_1)=0.6$ compared with the concentration inequality (CI) of Corollary \ref{cor: arma_is_concentrated} with 80\% confidence
 }
 \label{fig:histogram}
\end{figure}
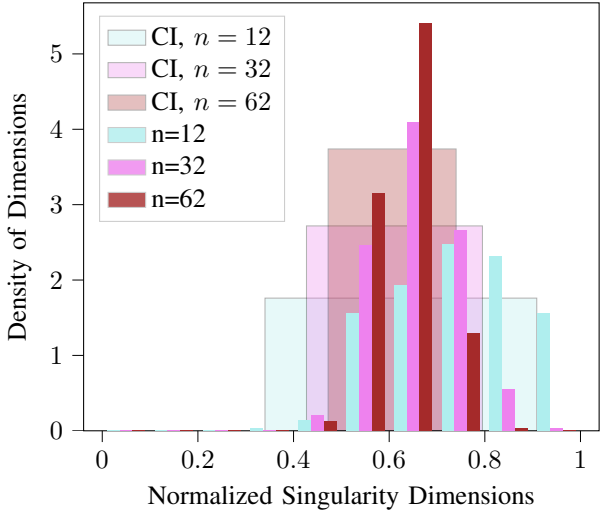

The latter result can be rephrased as an $(\epsilon, \delta)$ convergence criteria as in the following, showing that if the width of the truncation is large enough, then with probability at least $1-\epsilon$, the dimensions of singular components are concentrated within the interval $\big[n(\alpha-\delta), n(\alpha+\delta)\big]$.


\begin{Cor} \label{cor: arma_is_concentrated}
For a $(p,q)$-\ac{dce} process and given $\epsilon, \delta>0$, set 
\ea{
		n\geq \max \bigg\{\tfrac{2\big[q(1+\delta/2-\alpha)-p\big]}{\delta} &, \tfrac{-\log \frac{\ep}{2}}{D(\alpha-\delta/2\| \alpha)}-q ,\tfrac{2p}{\delta}\nonumber\\
		& , \tfrac{-\log \frac{\ep}{2}}{D(\alpha+\delta/2\| \alpha)}-q
		\bigg\},
	}
	and define $k_{-} = n(\alpha-\delta)$ and $k_{+} = n(\alpha+\delta)$. 
Theorem \ref{eqn: arma_is_affine} implies that $\Pr(d_i>k_{-})\geq 1-\frac{\ep}{2}$ and $\Pr(d_i<k_{+})\geq 1-\frac{\ep}{2}$, which in turn shows that
\ea{
		\Pr\big(\big|\tfrac{d_i}{n}-\alpha \big| < \delta\big) \geq 1-\ep.
	}

\end{Cor}

A representation of the concentration of measure in the above corollary is provided in Figure \ref{fig:histogram}.

To use such concentration property in showing our compressibility result, next we review a result that relates the BID of some sources to  their minimum $\ep$-achievable and Minkowski dimension compression rates.
\begin{Theorem}[\cite{CharusaieISIT2020P1}]\label{thm: concent_Rs}
Let $\{\Zv_t\}$ be a discrete-domain stochastic process such that the truncated sequence with $m$ samples (for all $n$) has an affinely singular distribution (Definition \ref{def:aff_sing}) with finitely many affine subsets $\{ \mathcal{A}_i\}_{i=1}^{k_n}$ with dimensions $\{d_i\}_{i=1}^{k_n}$ and corresponding measures $\{\mu_i\}_{i=1}^{k_n}$. Let $V_n$ be a discrete random variable that takes the values $1\leq i \leq k_n$ with probability $\mu_i(\mathcal{A}_i)$. If for all $\ep, \delta\in \Rbb^{+}$, there exists a large enough $n$ such that
\ea{
		\Pr\bigg(\Big|\tfrac{d_{V_n}}{m}-d_B\big(\{\Zv_t\}\big)\Big|<\delta\bigg)>1-\ep,	\label{eqn:Concentration}
	}  
	%
	then, for the process $\{\Zv_t\}$ (the typical source) we know that
	\ea{
		R^*(\ep)=R_B(\ep)=R(\ep)=d_B\big(\{\Zv_t\}\big),\label{eqn: Rs_equal}
	}
	where $R_B(\ep)$ is the Minkowski-dimension compression rate defined in \cite[Definition 10]{wu2010renyi}.
\end{Theorem}

Based on Theorem \ref{eqn: arma_is_affine} and Corollary \ref{cor: arma_is_concentrated}, we know that  \eqref{eqn:Concentration} holds for \ac{dce} processes with finite discrete space. Therefore, Theorem \ref{thm: concent_Rs} implies that \eqref{eqn: Rs_equal} is valid for such processes:
%
\begin{Theorem}\label{thm:comp_rate}
	Let the \ac{dce} process $\{\Zv_t\}$ as in Definition \ref{def: DC_ARMA} be such that
	\ea{
	 \xi_i = \nu_i X_{c, i} + (1-\nu_i) X_{D, i},
	}
where $\nu_i$s are i.i.d Bernoulli RVs with $\Pr(\nu_i = 1) = \alpha$, $X_{c, i}$s are absolutely continuous i.i.d RVs, and $X_{D, i}$s are discrete i.i.d RVs with $X_{D, i}\in \Dcal$ for a finite subset  $\Dcal$ of $\Rbb$. For the process $\{\Zv_t\}$ we have
\ea{
		R^*(\ep)=R_B(\ep)=R(\ep)= d(\xi_1) = \alpha.
}
\end{Theorem}

\begin{IEEEproof}
	See Appendix  \ref{app: ep_achievable_alpha}.
\end{IEEEproof}

We should highlight that in the case of discrete excitation noise ($\alpha=0$), the truncated process samples are no longer affinely singular, nevertheless, Theorem \ref{thm:comp_rate} still holds. 

\section{Conclusion}
\label{sec:Conclusion}

In this paper, we studied the compressibility of a class of ARMA processes from an information-theoretical perspective. 
More specifically, we considered the discrete-continuous DCE-ARMA processes, the excitation process of which have  discrete-continuous distributions. 
%
We showed that the discrete part of the excitation process induces certain types of singularity in the distribution of the samples path of the ARMA process, which greatly affects the overall compressibility. Besides evaluating various compressibility measures for these processes, we show that most of them (sample RID of the excitation process, minimum $\ep$-achievable compression rates of the ARMA process, BID of the ARMA process and the IDR of the ARMA process) are equal in this special case.
%


\bibliographystyle{IEEEtran}
\bibliography{arma}
\appendices

 \section{Continuity of ARMA samples}\label{app:ContARMA}
We first state the following lemma due to the clarity of our proofs.
\begin{lemma}\label{lem: sum_abs_cont}
	For an absolutely continuous RV $\Xv^n$ and an arbitrary RV $\Yv^n$, the sum  $\Zv^n= \Xv^n+\Yv^n$ is an absolutely continuous RV.
\end{lemma}
\begin{IEEEproof}
	See \cite[Theorem 2 and 4, Chapter V.4]{feller1971introduction}.
\end{IEEEproof}
\begin{Cor}\label{cor: sum_cont}
	Let $\Xv_1$ and $\Xv_2$ be independent $n$-dimensional RVs with  continuity chances (defined in Theorem \ref{thm: lrn}) $p_1$, $p_2$, respectively.
	The continuity chance $q$ of $\Zv=\Xv_1+\Xv_2$ satisfies
	\ea{
		q\geq 1-(1-p_1)(1-p_2)\geq \max\{p_1, p_2\}	. \label{eqn: just_a_cor}
	}
\end{Cor}

\begin{IEEEproof}
According to Lemma \ref{lem: sum_abs_cont}, we have that
\ea{
\mathbb{P} \big(\Zv \in \substack{\text{continuous}\\ \text{part}}\big) &\geq \mathbb{P} \Big(\big(\Xv_1 \in \substack{\text{continuous}\\ \text{part}}\big) \text{ or } \big(\Xv_2 \in \substack{\text{continuous}\\ \text{part}}\big)\Big) \nonumber\\
& \geq 1 - \underbrace{\mathbb{P} \big(\Xv_1 \not\in \substack{\text{continuous}\\ \text{part}}\big)}_{1-p_1}  \underbrace{\mathbb{P} \big(\Xv_2 \not\in \substack{\text{continuous}\\ \text{part}}\big)}_{1-p_2} \nonumber\\
& = 1-(1-p_1)(1-p_2)\geq \max\{p_1, p_2\},
}
where we used the independence of $\Xv_1$ and $\Xv_2$ to obtain the product of the probabilities.
%
%
\end{IEEEproof}


\begin{lemma}\label{lem: gen_conv}
Let $\Xv^{k_1}$ and $\Zv^{k_2}$ be absolutely continuous random vectors and $\Yv^{k_1}$ be a given random vector, and not necessarily absolutely continuous. Further, let $V$ be a random variable over $\Nbb$ such that given $V=i$, $\Xv^{k_1}$ is independent of $\Zv^{k_2}$ and $\Yv^{k_1}$. Then, for $\Scal\subseteq \mathbb{R}^{k_1+k_2}$, the probability measure 
$$\mu(\Scal)=\Pr\big([\Xv^{k_1}+\Yv^{k_1}; \Zv^{k_2}]\in \Scal \, \big| \, V=i \big)$$ is absolutely continuous for every $i\in \Nbb$.
%
\end{lemma}

\begin{rem}
If $\Xv^{k_1}$ is independent of $\Yv^{k_1},\, \Zv^{k_2}$ and $V=0$ with probability $1$, Lemma \ref{lem: gen_conv} implies that $[\Xv^{k_1}+\Yv^{k_1}; \Zv^{k_2}]$, and therefore $\Xv^{k_1}+\Yv^{k_1}$, are absolutely continuous. This shows that Lemma \ref{lem: sum_abs_cont} is a special case of Lemma \ref{lem: gen_conv}. Another special case is when $\Xv^{k_1},\, \Zv^{k_2}$ are independent, $\Yv^{k_1} \equiv \mathbf{0}$ and $V=0$ with probability $1$; in this case, Lemma \ref{lem: gen_conv} shows that concatenation of two independent absolutely continuous random vectors is again absolutely continuous.
%
\end{rem}

\begin{IEEEproof}
	We prove this lemma in two steps: (i) we show that the joint probability on a $(k_1\times k_2)$-dimensional box (cartesian product of closed intervals) can be rewritten in an integration form of a function $g(\av, \bv): \Rbb^{k_1}\times \Rbb^{k_2}\to \Rbb$, and (ii) we show that for any zero Lebesgue-measure set $\Scal$, the probability $\Pr\big([\Xv^{k_1}+\Yv^{k_1}; \Zv^{k_2}]\in \Scal\big |V=i)$ is zero.
%
	
		$\bullet$ {\bfseries Step (i)}:
		Since for two sets $\Rcal^{k_1}$ and $\Rcal^{k_2}$ in $\Rbb^{k_1}$ and $\Rbb^{k_2}$, we have 
		\begin{align}\label{eq:LHS_RHS}
		\Pr([\Xv^{k_1}+\Yv^{k_1}; \Zv^{k_2}] &\in \Rcal^{k_1}\times \Rcal^{k_2}|V=i) \nonumber\\
		&\leq \Pr(\Zv^{k_2}\in \Rcal^{k_2}|V=i),
		\end{align}
		 we can see that the LHS in \eqref{eq:LHS_RHS} is absolutely continuous w.r.t. the RHS (See \cite[Chapter V, 10.3]{feller1971introduction}). Hence, using Radon-Nikodym theorem, we can rewrite the LHS as 
		\ea{
			\Pr([\Xv^{k_1}+\Yv^{k_1}; &\Zv^{k_2}]\in \Rcal^{k_1}\times \Rcal^{k_2}|V=i)\nonumber\\& = \int_{\zv\in \Rcal^{k_2}} g( \Rcal^{k_1}, \zv) \Pr(\diff \zv|V=i), \label{eqn: joint_cond}
		} 
		where $g(\Rcal^{k_1}, \zv)$ is defined as
		\ea{
			g(\Rcal^{k_1}, \zv) = \lim_{h\to 0}	{ \Pr(\Xv^{k_1}+\Yv^{k_1}\in \Rcal^{k_1}|\Zv^{k_2}\in \Ical^{\zv}_h, V=i)},\label{eqn: cond}
		}
		in which $\Ical^{\zv}_h = [z_1, z_1+h]\times \ldots \times [z_{k_2}, z_{k_2}+h]$. This probability measure always exists (See \cite[Chapter V, 9.4]{feller1971introduction}). 
		
		Next, we rewrite the RHS of \eqref{eqn: cond} as follows
		\ea{
			&\Pr(\Xv^{k_1}+\Yv^{k_1}\in \Rcal^{k_1}|\Zv^{k_2}\in \Ical_{h}^{\zv}, V=i)\nonumber\\&= \int_{\yv\in \Rbb^{k_1}} \Pr(\Xv^{k_1}\in \Rcal^{k_1}-\yv |\Zv^{k_2}\in \Ical_{h}^{\zv}, V=i) \nonumber\\&\phantom{\int_{\xv\in \Rbb^{k_1}}}\times\Pr (\Yv^{k_1}\in \diff \yv|\Zv^{k_2}\in \Ical_{h}^{\zv}, V=i)\nonumber\\
			&\overset{(a)}{=} \int_{\yv\in \Rbb^{k_1}} \Pr(\Xv^{k_1}\in \Rcal^{k_1}-\yv| V=i) \nonumber\\&\phantom{\int_{\xv\in \Rbb^{k_1}}}\times\Pr (\Yv^{k_1}\in \diff \yv|\Zv^{k_2}\in \Ical_{h}^{\zv}, V=i),\label{eqn: conv_cond}
		}
		where $(a)$ is due to conditional independence of $\Xv^{k_1}$ and $\Zv^{k_2}$. Let $m(\Acal)$ stand for the Lebesgue measure of the Borel set $\Acal$. 
If  $\Pr(X\in \cdot| V=i)$ is an absolutely continuous measure, \cite[Prop. 15.5]{nielson1997introduction} implies that 
		for every $\ep>0$ there exists $\delta_{\ep}>0$ such that $\Pr(X\in \Acal|V=i)\leq\ep$ for every Borel set $\Acal\subset \Rbb^{k_1}$ with $m(\Acal)< \delta_{\ep}$.
		Further, using \cite[Theorem 12.1]{billing} we know that $m(\Acal) = m(\Acal-\yv)$ for every $\yv\in \Rbb^{k_1}$. Hence, for every choice of $\yv$ we have $\Pr(X\in \Acal-\yv|V=i)\leq\ep$. Thus, using  \eqref{eqn: conv_cond} we have that
		\ea{
			\Pr&(\Xv^{k_1}+\Yv^{k_1}\in \Acal|\Zv^{k_2}\in \Ical_{h}^{\zv}, V=i) \nonumber\\&\leq \ep \int_{\yv\in \Rbb^{k_1}} \Pr (\Yv^{k_1}\in \diff \yv|\Zv^{k_2}\in \Ical_{h}^{\zv}, V=i) \leq \ep.
		}
		This means that for every $\ep>0$, there exists $\delta_{\ep}>0$ such that if $m(\Acal)<\delta_{\ep}$ we can bound  $g(\Acal, \zv)$ as
		\ea{
			g(\Acal, \zv) = \lim_{h\to 0} \Pr&(\Xv^{k_1}+\Yv^{k_1}\in \Acal|\Zv^{k_2}\in \Ical_{h}^{\zv}, V=i) \leq \ep.	
			\label{eq:g_abs_cont}
		}
		Recall \cite[Prop. 15.5]{nielson1997introduction} , \eqref{eq:g_abs_cont} shows that 
		$g(\Acal, \zv)$ is an absolutely continuous measure. 
		
		Now, the Radon-Nikdoym theorem in conjunction with \eqref{eqn: joint_cond} reveals the existence of functions $q(\uv, \zv)$ and $q'(\zv)$ such that
		\ea{
			\Pr([\Xv^{k_1}+\Yv^{k_1}; &\Zv^{k_2}]\in \Rcal^{k_1}\times \Rcal^{k_2}|V=i) \nonumber\\&\overset{(a)}{=} \int_{\zv\in \Rcal^{k_2}} \int_{\uv\in \Rcal^{k_1}}q(\uv, \zv)\diff \uv \Pr(\diff \zv|V=i)\nonumber\\
			&\overset{(b)}{=} \int_{\zv\in \Rcal^{k_2}} \int_{\uv\in \Rcal^{k_1}}q(\uv, \zv)q'(\zv)\diff \uv\diff \zv,\nonumber
		}
		where $(a)$ is because of the absolute continuity of $g(\cdot, \zv)$ and $(b)$ holds due to the absolute continuity of $\Pr(\Zv^{k_2}\in \cdot|V=i)$. 
		
		$\bullet$ {\bfseries Step (ii)}:
		Let $\{\Rcal_k\}_{k=1}^{\infty}$ be an arbitrary countable set of (possibly intersecting) boxes. If we define $\Fcal_i=\Rcal_i \setminus \bigcup_{k=1}^{i-1}\Rcal_i$, we know that $\Fcal_i$s are disjoint:
\ea{ \label{eq:Fmeasure}
m\big( \cup_{i}\Rcal_i\big) = m\big( \cup_{i}\Fcal_i\big) = \sum_{i} m( \Fcal_i).
}		
		Besides, each  $\Fcal_i$ can be decomposed into finitely-many almost-disjoint boxes $\{\widetilde{\Rcal}_{i,j}\}_j$:
\ea{\label{eq:F_decomposition}
\exists \, \{  \widetilde{\Rcal}_{i,j} \}_{j=1}^{n_i}:~~ \left\{\begin{array}{l}
m\big(\widetilde{\Rcal}_{i,j_1} \cap \widetilde{\Rcal}_{i,j_2}\big) =0, ~~ j_1\neq j_2, \\ 
\Fcal_i \subseteq \cup_{j=1}^{n_i} \widetilde{\Rcal}_{i,j} , \phantom{\Big|}\\
m\Big(\big( \cup_{j=1}^{n_i} \widetilde{\Rcal}_{i,j} \big) \setminus \Fcal_i \Big) = 0. 
\end{array} \right.
}		
Note that $\widetilde{\Rcal}_{i,j}$s are closed boxes and  can overlap only at  their boundaries, i.e., $m\big(\widetilde{\Rcal}_{i,j_1} \cap \widetilde{\Rcal}_{i,j_2}\big) =0$. Similarly, $\Fcal_i$ and  $\cup_{j=1}^{n_i} \widetilde{\Rcal}_{i,j} $ might differ only at parts of the borders of some of  $\widetilde{\Rcal}_{i,j}$s. Overall, we have that
\ea{\label{eq:R_subset_Rtilde}
\big(\cup_i  \Rcal_i \big) = \big(\cup_i  \Fcal_i \big) \subseteq \big(\cup_{i,j} \widetilde{\Rcal}_{i,j} \big),
}
and
\ea{\label{eq:R_measure_equal_Rtilde}
m\Big( \big(\cup_{i,j} \widetilde{\Rcal}_{i,j} \big) \, \setminus \, \big(\cup_i  \Rcal_i \big) \Big) = 0.
}
%
In simple words, we showed that the union of an arbitrary countable set of boxes can be effectively decomposed into a union of countable set of almost-disjoint boxes.

		

		As a result, for every set of boxes $\{\Rcal_i\}_{i=1}^{\infty}$ we rewrite 
		\ea{
	&\Pr([\Xv^{k_1}+\Yv^{k_1}; \Zv^{k_2}]\in \cup_{k=1}^{\infty} \Rcal_k|V=i)\nonumber\\&\overset{(a)}{\leq }\Pr([\Xv^{k_1}+\Yv^{k_1}; \Zv^{k_2}]\in \cup_{i,j} \widetilde{\Rcal}_{i,j}|V=i)\nonumber\\	
	&\overset{(b)}{=}\sum_{i,j} \int_{(\uv, \zv)\in \widetilde{\Rcal}_{i,j}}   q(\uv, \zv)q'(\zv) \diff \uv \diff \zv\nonumber\\
	 &\overset{(c)}{=}\int_{(\uv, \zv)\in \cup_{i,j}\widetilde{\Rcal}_{i,j}} q(\uv, \zv)q'(\zv) \diff \uv \diff \zv\nonumber\\
	 &\overset{(d)}{=}\int_{(\uv, \zv)\in \cup_{k=1}^{\infty}{\Rcal}_k} q(\uv, \zv)q'(\zv) \diff \uv \diff \zv. \label{eqn: equal_prob}
}		
		Here, $(a)$ and $(d)$ are due to \eqref{eq:R_subset_Rtilde} and \eqref{eq:R_measure_equal_Rtilde}, respectively. 
		Further, $(b)$ is followed by Step (i), and $(c)$ is correct because $\widetilde{\Rcal}_j$s are almost-disjoint sets as \eqref{eq:F_decomposition} specifies.

		Now, we recall a result from \cite[Prop. 2.5.8]{athreya2006measure} that for every $L_1$-measurable function $f(\cdot)$ and constant $\ep>0$,
		there exists $\delta_{\ep}>0$ such that for every measurable set $\Acal$ with $m(\Acal)<\delta_{\ep}$, we have $\int_{\Acal} |f(\xv)|\diff \xv<\ep$. If we let $f(\uv, \zv) = q(\uv, \zv)q'(\zv)$,  one asserts that $f$ is an $L_1$ measurable function. The reason is that $\int_{\Rbb^{k_1+k_2}} f(\uv, \zv) \diff \uv \diff \zv=\Pr\big([\Xv^{k_1}+\Yv^{k_1}; \Zv^{k_2}]\in \Rbb^{k_1+k_2}|V=i\big) =1<\infty$. 
		
Let $\Scal$ be a zero Lebesgue-measure set. We choose $\ep>0$ arbitrarily small;	according to \cite[Prop. 2.5.8]{athreya2006measure} and \eqref{eqn: equal_prob}, there exists  $\delta_{\ep}$ such that if $m\big( \cup_{k=1}^{\infty}{\Rcal}_k \big) < \delta_{\ep}$, then,  $\Pr([\Xv^{k_1}+\Yv^{k_1}; \Zv^{k_2}]\in \cup_{k=1}^{\infty} \Rcal_k|V=i) < \ep$. Since  (see \cite[pp. 385, 389]{Pugh15})
\ea{
	\underbrace{m(\Scal)}_{=0} \triangleq 	\inf \Big\{\sum_k m(\Ocal_k): ~ \Ocal_k=\substack{\text{open} \\ \text{box}},\, \Scal\subseteq \cup_k \Rcal_k\Big\},
}		
we shall have a set of open boxes  	$\{\Ocal_k\}_k$ with $\sum_{k} m(\Ocal_k) < \delta_{\ep}$ such that $\Scal \subseteq \cup_k \Ocal_k$. If we set $\Rcal_k = \overline{\Ocal}_k$ (the closure of $\Ocal_k$), we have
\ea{
m\big(\cup_k \Rcal_k \big) \leq \sum_{k} m(\Rcal_k) = \sum_{k} m(\Ocal_k) <\delta_{\ep}.
}
In addition,
\ea{
\Scal \subseteq \cup_k \Ocal_k \subseteq \cup_k \Rcal_k.
}
If we summarize the above results, we achieve
\ea{ 
 & \Pr([\Xv^{k_1}+\Yv^{k_1}; \Zv^{k_2}]\in \Scal |V=i) \nonumber\\
& ~\leq 
\Pr([\Xv^{k_1}+\Yv^{k_1}; \Zv^{k_2}]\in \cup_{k=1}^{\infty} \Rcal_k|V=i) 
< \ep.
}
Since the choice of $\ep>0$ is arbitrary, we shall have $\Pr([\Xv^{k_1}+\Yv^{k_1}; \Zv^{k_2}]\in \Scal |V=i) =0$.

\end{IEEEproof}

\begin{lemma}\label{lem: hankel_full_rank}
Let $h[\cdot]$ be the causal inverse $z$-transform of a stationary ARMA filter $H(z)$ with $p+p'$ non-removable poles (see \eqref{eqn: transfer} for the definition of non-removable) from which $p$ are non-zero poles.
If for 
$i\geq \lceil \frac{p'+1}{p}\rceil + 1$ 
we define
	\ea{\label{eqn: Hankel}
	H^{(i)} = \Big[h[ip-p-1+j+k]\Big]_{j, k=0}^{p-1},
	} 
 then, $H^{(i)}$ is full-rank.
\end{lemma}


\begin{IEEEproof}
Let us define 
\begin{align}
\tilde{h}_i[n]= \left\{\begin{array}{lc}
h[ip-p-1+n], & n\geq 0,\\
0, & n<0.
\end{array} \right.
\end{align}
We first show that $\tilde{h}_i[\cdot]$ is the impulse response of a causal ARMA process with $p$ non-zero and non-removable poles:
\begin{align}
\tilde{H}_i(z) &= \sum_{n=0}^{\infty} \tilde{h}_i[n]z^{-n} = z^{ip-p-1} \sum_{n=ip-p-1}^{\infty} h[n]z^{-n} \nonumber\\
&= z^{ip-p-1} H(z) - \sum_{n=0}^{ip-p-2}h[n] z^{ip-p-1-n}. \label{eqn: poly}
\end{align}
Note that the last term of \eqref{eqn: poly} is polynomial in terms of $z$.
Since $i\geq \lceil \frac{p'+1}{p}\rceil + 1$ , we know that $ip-p-1 \geq p'$; therefore, $z^{ip-p-1} H(z)$ has no zero poles. This confirms that $\tilde{H}_i(z)$ has exactly $p$ non-removable poles all of which are non-zero. Thus, we can conclude the claim by recalling Lemma \ref{lem: hankel}.

We should highlight that since $\tilde{H}_i(z) = \sum_{n=0}^{\infty} \tilde{h}_i[n]z^{-n}$, we know that $\tilde{H}_i(z)$ is bounded as $z\to \infty$. Therefore, $\tilde{H}_i(z)$ is a \emph{proper} rational function.

\end{IEEEproof}

\begin{lemma}\label{lem: cond_abs_cont}
Let $\{\Xv_t\}$ be a D/C-ARMA process (see Definition \ref{def: DC_ARMA}) with excitation noise $\{\xiv_t\}$ and the corresponding RID of $d(\xi_1)\neq 0$. We know that a truncation $\xiv_{\Scal}=\{\xi_i:\, i\in \Scal\}$ of the excitation noise is affinely singular with a set of absolutely continuous RVs $\{\xitv_i\}$ laid on affine sets and a selection variable $V_{\Scal}$. The measure $\Pr([\Xv_{1}^{p}; \xitv_i]\in \cdot|V_{\Scal}=i)$ is absolutely continuous for all $i\in \Zbb$.
\end{lemma}
\begin{IEEEproof}
One could write $\Xv_1^p$ as 
	\ea{
		\Xv_1^{p} =  \sum_{i=0}^{\infty}\underbrace{H^{(i)} \boldsymbol{\xi}^{p-ip}_{1-ip}}_{\Yv_i},
		\label{eqn: T_xi}
	}
	where $H^{(i)}=\Big[h[ip-p-1+j+k]\Big]_{j, k=1}^{p}$, and $h[\cdot]$ is the causal inverse Z-transform of transfer function $H(z)$ of the D/C-ARMA process.
	Next, we rewrite $\Xv_{1}^{p}$ as
	\ea{
		\Xv_{1}^{p}=\underbrace{\sum_{i=0}^{i_o-1}\Yv_i}_{\Rv}+\underbrace{\sum_{i=i_0}^{\infty }\Yv_i }_{\Uv},
	}
	where $i_0=\max\big\{1+ \lfloor\frac{1-\min\Scal}{p} \rfloor,  1+\lfloor \frac{(q-p)_{+}+1}{p}\rfloor\big\}$. 
	
	To prove the theorem, we follow three steps: (i) we prove that  $\Uv$ is independent of $\Rv$ and $\xitv_i$ given $V_{\Scal}=i$, (ii) we show that $\Uv$ is an absolutely continuous RV, and (iii) we use Lemma \ref{lem: gen_conv} and the properties that are shown in the two previous steps to prove absolute continuity of $\Pr\big([\Rv+\Uv;\xitv_i]\in \cdot|V_{\Scal}=i\big)$.
 
		$\bullet$ {\bfseries Step (i)}: 
   Firstly, using \cite[Lemma 3
   ]{gyorgy1999rate}, we know that
   \ea{
   I(V_{\mathcal{S}} ; \xiv_{\mathcal{S}}) = H(V_{\mathcal{S}}),
   }
   or equivalently,
   \ea{
   H(V_{\Scal}|\xiv_{\Scal}) = 0.
   }
  As a result, for an arbitrary random variable $T$, and all sets $\Scal'$ that contain $\Scal$ we have
  \ea{
  I(V_{\Scal};T|\xiv_{\Scal'})\leq H(V_{\Scal}|\xiv_{\Scal'})\leq H(V_{\Scal}|\xiv_{\Scal}) = 0. \label{eqn: mutual_zero}
  }
  As a result, by assuming $\Scal''$ being mutually exclusive with $\Scal'$, we have
  \ea{
  I(\xiv_{\Scal''} ; \xiv_{\Scal'} | V_{\Scal}) &\leq I(\xiv_{\Scal''} ; \xiv_{\Scal'} , V_{\Scal})\\
  &= I(\xiv_{\Scal''} ; \xiv_{\Scal'})+I(\xiv_{\Scal''} ; V_{\Scal}| \xiv_{\Scal'}) = 0,
  }
  where the last equality holds because of \eqref{eqn: mutual_zero}, and mutually exclusiveness $\Scal'$ and $\Scal''$, and that $\{\xitv_t\}$ are drawn independently. 

  Moreover, because of chain rule we have
  \ea{
  I(\xiv_{\Scal''} ; \xiv_{\Scal'} | V_{\Scal})  = \sum_{i} \Pr(V_{\Scal} = i) I(\xiv_{\Scal''} ; \xiv_{\Scal'} | V_{\Scal} = i) = 0,
  }
  and since all the terms in RHS are non-negative, then for all $i$ such that $\Pr(V_{\Scal} = i)$ is non-zero, we have
  \ea{
  I(\xiv_{\Scal''} ; \xiv_{\Scal'} | V_{\Scal} = i) = 0. \label{eqn: mutual_given_each}
  }
  Moreover, we know that given $V_{\Scal}=i$, $\xitv_i$ is merely a projection of $\xiv_{\Scal}$ on an $e_i$-dimensional space where $e_i\leq |\Scal|$. As a result, given $V_{\Scal}=i$, $\xitv_{i}$ is a function of $\xiv_{\Scal}$. Hence, using data processing inequality we have
  \ean{
  0 \leq I(U; R, \xitv_i &|V_{\Scal} = i)\leq I(U; R, \xiv_{\Scal} | V_{\Scal} = i)\\
  & \overset{(a)}{\leq} I(\xiv_{-\infty}^{p-i_{0} p} ; R, \xiv_{\Scal} | V_{\Scal} = i)\\
  &\overset{(b)}{\leq} I(\xiv_{-\infty}^{p-i_{0} p}; \xiv_{p-i_0p+1}^{p}, \xiv_{\Scal} | V_{\Scal} = i ) \\
  &\overset{(c)}{\leq} I(\xiv_{-\infty}^{p-i_0p} ; \xiv_{p-i_0p+1}^{\max |\Scal|} | V_{\Scal} = i)\\
  &\overset{(d)} = 0,
  }
  where $(a)$ holds because $U$ is a function of $\xiv_{-\infty}^{p-i_0p}$ and because of data processing inequality, and $(b)$ holds because $R$ is a function of $\xiv_{p-i_0p+1}^{p}$ and as a result of data processing inequality. Moreover, $(c)$ is followed by $i_0\geq 1+ \big\lfloor\frac{1-\min \Scal}{p}\big\rfloor$ and as a result $\min \Scal \geq p-i_0p+1$. Finally $(d)$ is followed by \eqref{eqn: mutual_given_each}. This proves that $U$ and $R$ are independent given $V_{\Scal} = i$.

  $\bullet$ {\bfseries Step (ii)}: Firstly, since $d(\xi_1)=d>0$ and as a result of Lemma \ref{lem: cond_abs_cont}, the continuity chance of $\boldsymbol{\xi}_{1-ip}^{p-ip}$ is $d^p$.
	On the other hand, using Lemma \ref{lem: hankel_full_rank}, $H^{(i)}$ is a $p\times p$ full-rank matrix for $i\geq 3$. 
	Further, it is shown in \cite[Lemma 6]{CharusaieISIT2020P1} that the any full column-rank linear transformation of absolutely continuous random vectors is absolutely continuous.
	Hence, the continuity chance of $\Yv_i$ is at least $d^p$ for every $i\geq 3$.
	%
	Next, using the independence of $\Yv_i$s, and Corollary \ref{cor: sum_cont}, the continuity chance of $\Uv_t := \sum_{i=i_0}^{t} \Yv_i$ and also $\Uv$ is lower bounded by $1-(1-d^p)^{t-i_0}$. Finally, by arbitrariness of $t$, one can prove this claim.
 $\bullet$ {\bfseries Step (iii)}: We know by definition that $\tilde{\boldsymbol{\xi}}_i$ is an absolutely continuous random vector, and the absolute continuity of $\Uv$ is a result of Step (ii). Therefore, by making use of independence of $\Uv$ and $\Rv, \tilde{\boldsymbol{\xi}}_i$ given $V_{\Scal} = i$, and using Lemma \ref{lem: gen_conv} we conclude that $\Pr\big([\Rv+\Uv;\xitv_i]\in \cdot|V_{\Scal}=i\big)$ is an absolutely continuous measure that completes the proof.
\end{IEEEproof}

\section{Proof of Theorem \ref{eqn: arma_is_affine}}\label{app: arma_affine}
\begin{IEEEproof}
We prove this theorem using the following steps: (i) we show that a truncation $\xi_{p-q+1}^n$ of the excitation noise has an affinely singular probability distribution, (ii) using step (i) we show that the joint random vector $[\Xv^p; \xi_{p-q+1}^n]$ of a truncation of the ARMA process and its excitation noise is affinely singular, (iii) using the previous steps and linear recursive relation in ARMA processes we show that each truncation $\Xv^n$ of the ARMA process has affinely singular distribution, (iv) we show that if the excitation noise is absolutely continuous, then the truncation $\Xv^n$ is absolutely continuous, and (v) we use Lemma \cite[Lemma 7]{charusaie22} and Hankel property of the linear recursion of the ARMA process to show that the singularity dimensions of the truncation $\Xv^n$ concentrates around $n {\rm d}(\xi_i)$ for large $n$.

    $\bullet$ {\bfseries Step (i)}: We first should note that a truncation of discrete-continuous excitation noise $\{\xi_t\}$ of the \ac{dce} process has \emph{orthogonally singular} probability measure that is defined in \cite[Definition 2]{charusaie22}, which is known to be an affinely singular measure with singular components along the axes of the Euclidean space. To formalize this notion, assume that $\nu_t$ is a Bernoulli random variable that denotes whether $\xi_t$ takes its discrete ($\nu_t = 0$) or continuous values ($\nu_t = 1$).
    Using this choice of notation, \cite[Lemma 3]{charusaie22} shows that the truncation $\xi_{p-q+1}^n$ of the excitation noise has a probability distribution equivalent to
    \ea{
		\xiv^{n}_{p-q+1} \overset{d}{=}  U_V [\xitv_V ; 0_{d_O}]+\bv_V,	
	}
 where $V$ denotes the singular component on which the drawn instance of the random vector $\xiv^{n}_{p-q+1}$ is laid. Further, $U_V$ denotes a $(n+q-p)\times (n+q-p)$ permutation matrix that maps the first $d_V$ components of a vector to the indices of the components of $\xiv^{n}_{p-q+1}$ that take continuous values, i.e., $U_V = \big[I_{n+q-p}^{[\boldsymbol{\nu}]}, I_{n+q-p}^{[\overline{\boldsymbol{\nu}}]}\big]$, where $\boldsymbol{\nu}$ is, with an abuse of notation, defined as
 \ea{
  \boldsymbol{\nu} := \boldsymbol{\nu}_{p-q+1}^n.
 } 
Moreover, $\tilde{\xi}_V$ quantifies the distribution of the random vector $\xiv^{n}_{p-q+1}$ on its singular component with index $V$, $d_V$ is the Euclidean dimension of that singular component, and $d_{O}$ is the difference between the Euclidean dimension of that component and the encompassing space  $d_O = n+q-p-d_V$. 
Furthermore, $\bv_V$ is a fixed vector that determines the bias of the singular component from the origin.

$\bullet$ {\bf Step \bfseries (ii):} If we concatenate the truncation $\Xv^p$ of the process and the truncation $\xiv_{p-q+1}^n$ of the excitation noise, then by taking the first $d_V$ columns of $U_V$ and the $p$-dimensional identity matrix $I_p$ we can generate the tall matrix
\ea{
    \tilde{U}_V = \left[\begin{array}{c| c}
        I_p& 0_{p\times d_V}\\\hline
        0_{(n+q-p)\times p} & I_{n+q-p}^{\boldsymbol{[\nu]}}
    \end{array}\right],
}
using which the concatenation can be rewritten as
\ea{
		[\Xv^{p}; \xiv^{n}_{p-q+1}] =  \tilde{U}_V [\Xv^p; \xitv_{V}]+\bv_V.\label{eqn: joint_xs}
	}
Furthermore, Lemma \ref{lem: cond_abs_cont} shows that the concatenation $\Zv^{p+d_V} := [\Xv^p; \xitv_{V}]$ in RHS of the above identity is absolutely continuous for each choice of $V=i$. Therefore, if we complete the matrix $\tilde{U}_V$ by adding extra orthogonal columns, we can show that $[\Xv^{p}; \xiv^{n}_{p-q+1}]$ takes the form of an affinely singular random vector.

$\bullet$ {\bfseries Step (iii)}: In this step we use the concatenation $[\Xv^{p}; \xiv^{n}_{p-q+1}]$ that is shown to be affinely singular in previous step to specify the distribution of each truncation $\Xv^n$ of the ARMA process.  To that end, we refer the reader to the proof of Theorem \ref{thm: ARMA_inf_dim}, and particularly 
\eqref{eqn: theta_hat_x} where we show that $\Xv^n$ for $n\geq p$ can be written in form of
\ea{
\Xv^n = \widehat{\Phi}^{-1} \widehat{\Theta} \big[\Xv^p; \xi_{p-q+1}^n\big], \label{eqn: xn_interms_joint}
}
where $\widehat{\Phi}$ and $\widehat{\Theta}$ are defined in \eqref{eqn: phi_hat_def} and \eqref{eqn: theta_hat_def}, respectively. Using \eqref{eqn: joint_xs}, the above identity can be rewritten as
\ea{
\Xv^n =  \widehat{\Phi}^{-1} \widehat{\Theta}\tilde{U}_V [\Xv^p; \xitv_{V}]+\bv'_V, \label{eqn: phihat_xn}
}
where $\bv'_V = \widehat{\Phi}^{-1} \widehat{\Theta}$ is a fixed vector for each choice of $V$. Therefore, in order to show that $\Xv^n$ has an affinely singular distribution, we need to rewrite the first term of RHS of \eqref{eqn: phihat_xn} as a rotation of a low-dimensional absolutely continuous random vector. To that end, we first rewrite the matrix $\widehat{\Phi}^{-1} \widehat{\Theta}\tilde{U}_V$ using the method of singular value decomposition (SVD) as $ \widehat{\Phi}^{-1} \widehat{\Theta}\tilde{U}_V = Q_V \Sigma_V P_V^{\dagger}$, where $Q_V$ and $P_V$ are unitary matrices and $\Sigma_V$ is a diagonal matrix with ${\rm rank}(\widehat{\Phi}^{-1} \widehat{\Theta}\tilde{U}_V)$ non-zero terms. Next, we use \cite[Lemma 10]{charusaie22} that shows that the result of a multiplication of a full-row rank matrix and an absolutely continuous RV is another absolutely continuous RV. In fact, $\Sigma_V P_V^{\dagger}$ is a concatenation of a full-row rank matrix and null rows. Therefore, using absolute continuity of $\big[\Xv^p; \xi_{p-q+1}^n\big]$ in Step (ii), we conclude that $\Sigma_V P_V^{\dagger}\big[\Xv^p; \xi_{p-q+1}^n\big]$ is a concatenation of an absolutely continuous RV with dimension ${\rm rank}(\widehat{\Phi}^{-1} \widehat{\Theta}\tilde{U}_V)$ and zero components. This, together with the SVD form shows that $\Xv^n$ is an affinely singular RV in which the singular components have dimension $d_V = {\rm rank}(\widehat{\Phi}^{-1} \widehat{\Theta}\tilde{U}_V)$.

By definition of $\widehat{\Phi}$ in \eqref{eqn: phi_hat_def}, we know that it is a full-rank square matrix. Therefore, we have
\ea{
 {\rm rank}(\widehat{\Phi}^{-1} \widehat{\Theta}\tilde{U}_V) = {\rm rank}(\widehat{\Theta}\tilde{U}_V).
}
Moreover, by definition of $\widehat{\theta}$ in \eqref{eqn: theta_hat_def}, we can show that
\ea{
\Theh \Ut_V = 	\left[\begin{array}{c| c}
			I_p & 0_{p\times d_V}\\\hline
			0_{n-p\times p} & \Theta^{[\boldsymbol{\nu}]}
		\end{array}\right],
}
that has the rank of $p+{\rm rank}(\Theta^{[\nuv]})$. Therefore, using the above two identities, we have
\ea{
d_V = p + {\rm rank}(\Theta^{[\nuv]}). \label{eqn: dv_def}
}

$\bullet$ {\bfseries Step (iv)}: In case of absolute continuity of the excitation noise, and the definition of $\nuv$ in Step (i) assures that $\nuv$ is an all-one vector. Therefore, $d_V$ in \eqref{eqn: dv_def} can be obtained as
\ea{
d_V = p + {\rm rank}(\Theta),
}
and since $\Theta$ is a full-row rank matrix with $n-p$ rows (see the definition of $\Theta$ in \eqref{eq:theta}), therefore we have
\ea{
d_V = n,
}
that is equivalent to say $\Xv^n$ is a mixture of absolutely continuous RVs, and therefore itself is an absolutely continuous RV.

$\bullet$ {\bfseries Step (v)}: In case that the excitation noise is discrete-continuous with continuity chance of $\alpha$, then the random vector $\nuv$ is distributed as $n-p+q-1$ i.i.d Bernoulli random variables with each components having $\Pr(\nu_i=1)=\alpha$. As a result, the matrix $\Theta^{[\nuv]}$ is a random choice of columns of a Hankel matrix with probability $\alpha$. For such random choice, \cite[Lemma 8]{charusaie22} shows that the rank function is distributed as
\ea{
		\Pr \big(\rank(&\Theta^{[\boldsymbol{\nu}]})> k\big)\nonumber\\&\geq 1-\exp\Big(-(n+q-p)D\big(\tfrac{k+q}{n+q-p}\|\alpha\big)\Big),
	}
	for $\tfrac{k+q}{n+q-p}<\alpha$ and 
	\ea{
		\Pr \big(\rank(\Theta^{[\boldsymbol{\nu}]})&< k\big)\nonumber\\&\geq 1-\exp\Big(-(n+q-p)D\big(\tfrac{k}{n-p}\|\alpha\big)\Big),
	}
	for $\tfrac{k}{m}>\alpha$. These inequalities together with \eqref{eqn: dv_def} conclude in \eqref{eqn: conc_more} and \eqref{eqn: conc_less}, and accordingly complete the proof.

	for $\tfrac{k}{m}>\alpha$. Using these inequalities and the above discussions, one can directly derive \eqref{eqn: aff_arma}.

\end{IEEEproof}

\section{Proof of Theorem \ref{thm: ARMA_inf_dim}}\label{app: ARMA_inf_dim}
	Before delving into the details of the proof, we introduce the following lemma to bound the entropy of quantized shifted version of a RV.
\begin{lemma} \label{lem: bounded_shifted}
	For a RV $X$, if we have $H\big([X]_1\big)<m$, then, we can bound $H\big([X+\ep]_1\big)$ as
	\ea{
		H\big([X+\ep]_1\big)\leq 4m+{\rm C},
	}
	for every $\ep\in (-1, 1)$, where $\rm C$ is a fixed value that dependent on the distribution of $X$, and not $\ep$.
\end{lemma}

\begin{IEEEproof}
	We define $Y=X+\ep$, and study the entropy
	\ea{
	H\big([Y]_1\big) = -\sum_{i=-\infty}^{\infty} p_i^Y \log p_i^Y,
	}
where $p_i^Y$ is defined as
\ea{
p_i^Y = \Pr(i\leq Y<i+1).
}
Let $c\in \Nbb$ be a large enough integer such that $\Pr(|X|>c)<1/4$. We now
decompose $H\big([Y]_1\big)$ as
\ea{
H\big([Y]_1\big) = &-\sum_{i\in [-c-1:c]} p_i^Y \log p_i^Y \nonumber\\&- \sum_{i\geq c+1, i\leq -c-2} p_i^Y \log p_i^Y. \label{eqn: decompose_H_y}
}
By defining  
\begin{align}
A=\sum_{i\in [-c-1:c]} p_i^Y &= \Pr(-c-1\leq Y<c+1) \nonumber\\
&\geq \Pr(|X|<c) \geq \tfrac{3}{4},
\end{align}
we can write
\ea{
-\sum_{i\in [-c-1:c]}\hspace{-.5cm} p_i^Y \log p_i^Y &= -A\hspace{-.5cm} \sum_{i\in [-c-1:c]} \underbrace{\frac{p_i^Y}{A}}_{\tilde{p}_i}\log \frac{p_i^Y}{A} A\\
& = - A\hspace{-.5cm}\sum_{i\in [-c-1:c]}\hspace{-.5cm} \pt_i \log \pt_i - A\log A\\
&\overset{(a)}{\leq} A (-\hspace{-.5cm}\sum_{i\in [-c-1:c]}\hspace{-.5cm} \pt_i \log \pt_i) +\tfrac{1}{3}\\
&\overset{(b)}{\leq} \log \big(2(c+1)\big) +\tfrac{1}{3}, \label{eqn: upper_bounded}
}
where $(a)$ is followed by $A\in[ \frac{3}{4}, 1]$, and $(b)$ is because the entropy of any discrete RV with $2(c+1)$ symbols is upper-bounded by $\log\big(2(c+1)\big)$

Let $s_{\ep}=1$ if $\ep \in [0, 1)$, and $s_{\ep}=-1$ if $\ep \in (-1, 0)$. We know that for all $i\geq c+1, i\leq -c-2$ 
\ea{
p_i^Y \leq p_i^X+p_{i+s_{\ep}}^X \leq \sum_{k\geq c+1, k\leq -c-2} p_k^X \leq \tfrac{1}{4}.
}
Since $-x\log x$ is an increasing function for $x\leq \frac{1}{2}$, we conclude that
\ea{
-p_i^Y &\log p_i^Y  \leq -(p_i^X+p_{i+s_{\ep}}^X) \log (p_i^{X}+p_{i+s_{\ep}}^X) \nonumber\\
&\leq -2\max\{p_i^X, p_{i+s_{\ep}}^X\}\log \big(2\max\{p_i^X, p_{i+s_{\ep}}^X\} \big) \nonumber\\
&\overset{(a)}{\leq} - 2p_i^X \log \big(2p_i^X\big) - 2p_{i+s_{\ep}}^X \log \big(2p_{i+s_{\ep}}^X\big) \nonumber\\
&\leq -2p_i^X \log p_i^X - 2p_{i+s_{\ep}}^X \log p_{i+s_{\ep}}^X - 2(p_i^X +p_{i+s_{\ep}}^X )
\label{eqn: new_bound_term_xlogx} 
}
%
%
where $(a)$ is followed by positiveness of function $-x\log x$ for $x\leq 1$ coupled with the fact that $p_i^X, p_{i+1}^X\leq \frac{1}{4}$. 

Finally, using \eqref{eqn: decompose_H_y}, \eqref{eqn: upper_bounded}, and \eqref{eqn: new_bound_term_xlogx} we have
\ea{
H\big([Y]_1\big)&\leq \log (c+1) +\frac{4}{3} + 4H\big([X]_1\big)-4\\&\leq 4m+\underbrace{\log (c+1)-\tfrac{8}{3}}_{\rm C}.
}
%
\end{IEEEproof}

%

\begin{IEEEproof} [Proof of Theorem \ref{thm: ARMA_inf_dim}]
%
	%
	Consider the ARMA process at time $t>0$ and the corresponding $\Phi$ and $\Theta$ matrices. We first modify $\Phi$  to make it invertible:
	\ea{\label{eqn: phi_hat_def}
	\widehat{\Phi} &= \left[\begin{array}{c}
			\begin{array}{c| c}
				I_{p} & 0_{p\times m}
			\end{array}\\\hline
			\Phi_{m\times (m+p)}
		\end{array}\right].
	}
	We can now rewrite \eqref{eq:vector expression} as
	\ea{
		\Phih \cdot\Xv^{m+p} & = \Theh \underbrace{ \lsb \p { \Xv^{p} \\ \xiv_{1-q+p}^{m+p} } \rsb }_{\Qhv} \label{eqn: theta_hat_x} =  \Theh \Qhv,
	}
	where
	\ea{
		\widehat{\Theta} =\left[
		\begin{array}{c| c}
			I_{p} & 0_{p\times(m+q)}\\\hline
			0_{m\times p} & \Theta_{m\times (m+q)}
		\end{array}
		\right]\label{eqn: theta_hat_def}.
	}
	%
	%
	From the fact that $\det(\Phih)=1$ and \cite[Thm. 2]{WuThesis}, it follows that
	\ea{
		d(\Xv_{1}^{m+p})=d(\Theh \Qhv). \label{eq:d_equality}
	}
	The remainder of the proof relies on the following steps: (i) we show that $[\Xv^p, \xiv^p_{1-q+p}]$ is independent of $\xiv_{p+1}^{p+m}$,  (ii) we find  lower- and upper-bounds for $d(\Xv^{m+p})$ and determine $d_{B}\big(\{\Xv_t\}\big)$, and finally, (iii) we prove that $d_B\big(\{\Xv_t\}\big)$ is a lower-bound for $d_I\big(\{\Xv_t \}\big)$. The latter bound together with the result in Theorem \ref{th:koch2019} completes the proof.
	
	{$\bullet$ \bfseries Step (i)}:
	Using the recurrence relationship in \eqref{eqn: arma_eqn}, we know that $\Xv^p$ is a function of excitation noise samples $\xiv_{-\infty}^{p}$. Since $\{\xi_t\}$ is an i.i.d. process, it is straightforward to check that $[\Xv^p, \xiv^p_{1-q+p}]$ is again a function of $\xiv_{-\infty}^{p}$ and independent of $\xiv_{p+1}^{p+m}$.
	
%
	
	{$\bullet$ \bfseries Step (ii)}:
	\vspace{-1mm}
	As $\widehat{\The}$ in \eqref{eq:d_equality}
	forms a linear, and therefore Lipschitz,  transform, we know from \cite[Thm. 2]{WuThesis} that
	\eas{
		d(\Xv^{m+p}) = d(\Theh \Qhv)&\leq d(\Xv^p, \xiv_{1-q+p}^p, \xiv_{p+1}^{p+m}) \nonumber \\
		& =d(\Xv^p, \xiv_{1-q+p}^p) +d(\xiv_{p+1}^{p+m})\label{eqn: d_x_m_upper_1} \\
		& \leq p+q+m\alpha,\label{eqn: d_x_m_upper_2}
	}{\label{eqn: d_x_m_upper}}
	where \eqref{eqn: d_x_m_upper_1} is due to the independence of $(\Xv^p, \xiv_{1-q+p}^p)$ and $\xiv_{p+1}^{p+m}$  followed by \cite[Eqn. 2.22]{WuThesis},
	while \eqref{eqn: d_x_m_upper_2} is due to \cite[Eq. 2.11]{WuThesis} because of the independence of the elements of $\{\xiv_t\}$. 
	Besides, we can partition $\Theh\Qhv$ as follows,
	\ea{
		\widehat{\The}\Qhv=
		\lsb\p{
			\widehat{\The}_{1}
			& \rvline & \widehat{\The}_{2} &\rvline &  \widehat{\The}_{3} \\
		}\rsb
		\cdot
		\lsb \p { \Xv^p \\  \xiv^p_{1-q+p} \\ \xiv_{p+1}^{m+p}   } \rsb,
	}
	so that we can define
	\eas{
		U_1  &= \widehat{\The}_{1}  \Xv^p+ \widehat{\The}_2  \xiv^p_{1-q+p} ,\\
		U_2 &= \widehat{\The}_3 \xiv_{p+1}^{m+p}.
	}
	Since $\xiv_{p+1}^{p+m}$ is independent of $\Xv^{p}$ and $\xiv_{1-q+p}^p$ jointly, we conclude that  $U_1$ and $U_2$ are also independent.
	%
	%
	%
	%
	Therefore, using \cite[Eqn. 20.20]{WuThesis}, we have that
	\ea{
		d(\Theh\Qhv) = d(U_1+U_2)\geq d(U_2). \label{eqn: up_d_u2}
	}
	We recall that $\widehat{\The}_3$ is a lower-triangular square matrix with all the diagonal elements being $1$. Hence (based on \cite[Thm. 2]{WuThesis}),
	\ea{
	d(U_2) = d\big( \widehat{\The}_3 \xiv_{p+1}^{m+p} \big) = d\big( \xiv_{p+1}^{m+p} \big) = m\alpha.
	}
	Overall, this implies that
	\ea{
		d(\Xv^{p+m})=d(\Theh\Qhv) \geq  m \alpha. \label{eqn: d_x_m_lower}
	}
%
Combining \eqref{eqn: d_x_m_upper} and \eqref{eqn: d_x_m_lower}, we can write
\ea{
\frac{m\alpha }{m+p} \leq \f {d(\Xv^{p+m})}{m+p} \leq \frac{p+q+m\alpha }{ m+p},
}
which proves that
\ea{
		d_B\big(\{\Xv_m\}\big) = \lim_{ m \to\infty} \f {d(\Xv^{p+m})}{m+p} = \alpha. \label{eq:d_B_equal _alpha}
	}

	
	{$\bullet$ \bfseries Step (iii)}:
	Using \cite[Eqn. 16, 73]{koch2019}, we write that
	\ea{
		d_I\big(\{\Xv_t\}\big)	\geq d(X_{p+1}|\Xv_{-\infty}^p, \xiv_{1-q+p}^{p}).\label{eqn: d_I_lower}
	}
According to the definition of RID, we know	
\ea{
		d(X_{p+1}|&\Xv_{-\infty}^p, \xiv_{1-q+p}^{p}) 
		= \lim_{k\to\infty} \f{H\big([X_{p+1}]_k|\Xv_{-\infty}^p, \xiv_{1-q+p}^{p}\big)}{\log k} \nonumber\\
		&\phantom{\Xv_{-\infty}^p, \xiv_{1-q+p}^{p})} \overset{(a)}{=}  \lim_{k\to\infty} \f{H\big([X_{p+1}]_k|\Xv^p, \xiv_{1-q+p}^{p}\big)}{\log k} \nonumber\\
		&=	\lim_{k\to\infty}\int_{\Rbb^{p}\times \Rbb^q} \f {\Hcal_k(\xv^p, \boldsymbol{\tau}_{1-q+p}^p)}{\log k} \diff \mu(\xv^p, \boldsymbol{\tau}_{1-q+p}^p),\label{eqn: int_cond_inf_dim_2}
	}
	where $(a)$ is because of \eqref{eqn: arma_eqn}, 
%
	$\mu(\cdot, \cdot)$ is the joint probability measure on $(\Xv^p, \xiv_{1-q+p}^{p})$ and
	\ea{
		\Hcal_k(\xv^p, \boldsymbol{\tau}_{1-q+p}^p) = H\big([X_{p+1}]_k|\Xv^p = \xv^p,\xiv_{1-q+p}^{p} = \boldsymbol{\tau}^{p}_{1-q+p} \big).
	}
	Combining \eqref{eqn: arma_eqn} and \eqref{eqn: int_cond_inf_dim_2} we have that
	\ea{
		\Hcal_k(\xv^p, \boldsymbol{\tau}_{1-q+p}^p) =  H\big([\xi_{p+1}+c]_k\big)\label{eqn: cond_dith_2},
	}
	where
	\ea{
		c = \sum_{i=1}^{q}\theta_i\tau_{p+1-i}+\sum_{i=1}^{p} \phi_ix_{p+1-i}.
	}
	
	To further simplify \eqref{eqn: cond_dith_2}
	\ea{
	H\big([\xi_{p+1}+c]_k\big) &= H\big( \big[\xi_{p+1}+c - [c]_k\big]_k \big) \nonumber\\
	& \overset{(a)}{\leq } H\big( \big[\xi_{p+1}+c - [c]_k\big]_1 \big) + \log(k),
	}
	where $(a)$ is valid because of \cite[Eqn. 11]{renyi1959dimension}. Consequently, for $k\geq 2$, we have 
	\ea{
		\f{H\big([\xi_{p+1}+c]_k\big)}{\log k} \leq \f{ H\Big(\big[\xi_{p+1}+c-[c]_k\big]_1\Big)}{\log k}+1.
	}
	Since $c-[c]_k \in [0, 1)$ and $H\big([\xi_{p+1}]_1\big)\leq m<\infty$, we can apply Lemma \ref{lem: bounded_shifted} to conclude $H\big([\xi_{p+1}+c-[c]_k]_1\big)<4m+{\rm C}$, where $\rm C$ is a fixed value and merely dependents on the distribution of $\xi_{p+1}$. As a result, we have that
	\ea{
		\f{H\big([\xi_{p+1}+c]_k\big)}{\log k}\leq \f{4m+{\rm C}}{\log k}+1.\label{eqn: bound_renyi_xi_2}
	}
	By plugging in the latter result and \eqref{eqn: cond_dith_2} in \eqref{eqn: int_cond_inf_dim_2}, we get
	\ea{
	&\int_{\Rbb^{p}\times \Rbb^q} \f {\Hcal_k(\xv^p, \boldsymbol{\tau}_{1-q+p}^p)}{\log k} \diff \mu(\xv^p, \boldsymbol{\tau}_{1-q+p}^p) \nonumber\\
	&\leq \Big( \f{4m+{\rm C}}{\log k}+1 \Big) \underbrace{\int_{\Rbb^{p}\times \Rbb^q} \diff \mu(\xv^p, \boldsymbol{\tau}_{1-q+p}^p)}_{1}  \leq \f{4m+{\rm C}}{\log 2}+1.
	}
	This allows us to apply the dominant convergence theorem to achieve
	\ea{
	d(X_{p+1}|&\Xv_{-\infty}^p, \xiv_{1-q+p}^{p}) \nonumber\\
	&= \int_{\Rbb^{p}\times \Rbb^q} \Big(\lim_{k\to\infty} \f {\Hcal_k(\xv^p, \boldsymbol{\tau}_{1-q+p}^p)}{\log k} \Big) \diff \mu(\xv^p, \boldsymbol{\tau}_{1-q+p}^p) \nonumber\\
	&= \int_{ \Rbb^{p}\times \Rbb^q} d\Big(\xi_{p+1}+\sum_{i=1}^{q}\theta_i\tau_{p+1-i}+\sum_{i=1}^{p} \phi_ix_{p+1-i}\Big)\nonumber\\&\qquad\qquad\qquad\times\diff \mu(\xv^p, \boldsymbol{\tau}_{1-q+p}^p ).
	}
%
\cite[Lem. 3]{WuThesis} implies that 
\ea{
d\Big(\xi_{p+1}+\sum_{i=1}^{q}\theta_i\tau_{p+1-i}+\sum_{i=1}^{p} \phi_ix_{p+1-i}\Big)
= d(\xi_{p+1}) = \alpha,
}
which means
	\ea{
		d(X_{p+1}|\Xv_{-\infty}^p, \xiv^{p}) = \alpha.
	}
	Using this identity and \eqref{eqn: d_I_lower} one can see that
	\ea{
		d_I\big(\{\Xv_t\}\big)\geq \alpha.
		\label{eq:inequality alpha}
	}
We recall from 	Theorem \ref{th:koch2019} that $d_{I}(\{ \Xv_t\}) \leq d_{B}(\{ \Xv_t\})$. However, \eqref{eq:d_B_equal _alpha} states that $d_{B}(\{ \Xv_t\}) =\alpha$. Now, \eqref{eq:inequality alpha} shows that $d_{I}(\{ \Xv_t\}) = d_{B}(\{ \Xv_t\}) = \alpha$.
\end{IEEEproof}

\section{Proof of Theorem \ref{thm:comp_rate}}\label{app: ep_achievable_alpha}
	\begin{IEEEproof}
		We consider the three cases of discrete, continuous, and discrete-continuous excitation noise separately.
		
			$\bullet$ {\bfseries Discrete excitation ($\Pr(\nu_i=1)=0$)}:  In this case, we are dealing with infinite mixtures of purely discrete RVs; therefore, we no longer have affinely singular random variables (there is no absolutely continuous component even on lower dimensional subspaces). In this case, we show that
			$\ep$-achievable rates are zero. This together with the fact that $d(\xi_i)=0$ for a discrete random variable $\xi_i$ (see Theorem \ref{thm:disc_cont_dim}), proves the  theorem. 
			
			Using the definition of ARMA process in  \eqref{eqn: arma_eqn}, we know that 
			\ea{
			X_t+\sum_{i=1}^p \phi_i X_{t-i} = \xi_t +\sum_{i=1}^q \theta_i\xi_{t-i}.
			}
			Since besides the excitation noise, each $X_t$ depends on its $p$ previous samples, the recursive equation is solvable by having $p$ boundary conditions (the the realization of the excitation noise).
			As a result, $\Xv^n$ could be expressed as a linear function of $\Xv^p$ and $\xi_{1-q}^n$ for each $n$.
            Since $\xi_i$ takes value from a finite or countably infinite set, then we can find a finite set $\Acal_i$ such that
            \ea{
            \Pr(\xi_i \in \Acal_i) \geq 1 - \frac{\delta}{2^{i+q}},
            }
            for all $i\in \{1-q , \ldots\}$, where $\delta\in(0, 1)$. Therefore, the probability of $\xi_{1-q}^n$ taking a value from $\Bcal_n = \Acal_i \otimes \ldots \otimes \Acal_{q+n}$ is lower-bounded as 
            \ea{
                \Pr (\xi_{1-q}^n\in \Bcal_n) \geq \prod_{i=1}^{n+q} \big(1-\frac{\delta}{2^i}\big) \geq 1-\delta\sum_{i=1}^{n+q} \frac{1}{2^i} \geq 1-\delta. \label{eqn: choose_finite}
            }
            Next, given $\xi_{1-q}^n\in \Bcal_n$, since $\Xv^n$ is obtained via a linear function $\Xv^n = f(\Xv^p, \xi_{1-q}^n)$ of the excitation noise and previous samples, and because of finiteness of $\Bcal_n$ we can cover the support of $\Xv^n$ with a finite set of linear functions of $\Xv^p$. More formally, if we define a random vector $\Yv^n$ that is distributed as
            \ea{
			\Pr(\Yv^n = \yv^n)=\Pr(\Xv^n = \yv^n|\xiv_{1-q}^n \in \Acal_m^n), \label{eqn: dist_y}
			}
            then there exists a set of functions $f_i:\Rbb^p \to \Rbb^n$ for $i\in [1:|\Bcal_n|]$ such that $\mathrm{supp}(\Yv^n)\subseteq \cup_{i=1}^{n} \mathrm{supp}\big(f_i(\Xv^p)\big)$. A set with such property (i.e., covered by a finite number of Lipschitz images of $\Rbb^p$) is called $p$-rectifiable.  We observe that rectifiability is further satisfied for all higher values of $p$, i.e. we can cover the support of $\Yv^n$ by adding dummy dimensions to $\Xv^p$. As stated in \cite[Lemma 12]{wu2010renyi}, if $\supp(\Yv^n)$ is $\lfloor Rn\rfloor$-rectifiable for all sufficiently large $n$s, then the minimum $\ep$-achievable rate for $\{\Yv_i\}$ is bounded above by $R$. Hence, this property is met for $\Yv^n$, and $R=\frac{p}{m}$ and all $n\in \{m, m+1, \ldots\}$. As a result, $R_{\{\Yv_i\}}(\ep)\leq \frac{p}{m}$. 

            Next, to find an upper-bound for minimum $\ep$-achievable rate for $\{\Xv_i\}$ by $R'$, one needs to find a set of $(n, \lfloor R'n\rfloor)$-encoder-decoder pair $g_i, h_i$ such that
			\ea{
			\Pr(g_n(h_n(\Xv^n))\neq \Xv^n)\leq \ep.
			}
			Here, if we choose the same pair of encoders and decoders as of $\{\Yv_i\}$, we have
			\ea{
			\Pr&(g_n(h_n(\Xv^n))\neq \Xv^n) \nonumber\\&= \Pr(g_n(h_n(\Xv^n))\neq \Xv^n|\xiv_{1-q}^n \in \Acal_m^n)\Pr(\xiv_{1-q}^n \in \Acal_m^n) \nonumber\\&+ \Pr(g_n(h_n(\Xv^n))\neq \Xv^n|\xiv_{1-q}^n \notin \Acal_m^n)\Pr(\xiv_{1-q}^n \notin \Acal_m^n),
			}
			that coupled with \eqref{eqn: choose_finite} and \eqref{eqn: dist_y} concludes in
			\ea{
			\Pr(g_n(h_n(\Xv^n))\neq \Xv^n)&\leq \Pr(g_n(h_n(\Yv^n))\neq \Yv^n) +\frac{1}{m}\\
			&\leq \ep' +\delta.
			}
			By having large enough $n$ such that $\ep'\leq \frac{\ep}{2}$ and by setting $\delta \in (0, \tfrac{\ep}{2})$, we have $R(\ep)\leq \frac{p}{m}$. Since we could fix $m$ arbitrarily large, and by positiveness of $R(\ep)$, we can prove that
			\ea{
			R(\ep)=0.
			}
			The above identity together with the inequality
			\ea{
			R^*(\ep)\leq R_{B}(\ep)\leq R(\ep),
			}
			in \cite[Eq. 75]{wu2010renyi} proves that all above values of $R*(\ep), R_B(\ep), R(\ep)$ are equal to $0$.

			$\bullet$ {\bfseries Continuous excitation ($\Pr(\nu_i=1)=1$)}:
			%
			In this case, the first part of Theorem  \ref{eqn: arma_is_affine} shows that for all $n$, $\Xv^n$ is also absolutely continuous. 
			Now, Theorem \ref{thm:disc_cont_dim} implies that $d(\Xv^n) = n$. Besides,  Theorem \ref{thm:BID sss} reveals that $d_B\big(\{\Xv_t\}\big)= \lim_{n\to\infty} \frac{d(\Xv^n)}{n}=1$. 
			
			Due to absolute continuity of $\Xv^n$, it can be thought of as an affinely singular random variable 
			$V_n=1$ and $d_1=n$. As a result, one can see that $\Big|\frac{d_1}{n}-d_{B}\big(\{\Zv_t\}\big)\Big|=0$. Equivalently, we have
			\ea{
			 \Pr_{V_n}\bigg(i: \Big|\tfrac{d_i}{n}-d_B\big(\{\Zv_t\}\big)\Big|<\delta\bigg)=1>1-\ep,
			}
			for every pair of $(\delta, \ep)$. Finally, using Theorem \ref{thm: concent_Rs}, we conclude that 
			\ea{
				R^*(\ep)=R_B(\ep)=R(\ep)=d_B\big(\{\Zv_t\}\big)=d(\xi_1)=1,	
			}
			where the last equality holds  because of  Theorem \ref{thm:disc_cont_dim}.
		
			$\bullet$ {\bfseries Mixed discrete-continuous excitation ($\Pr(\nu_i=1)\in (0, 1)$)}:
			Recalling Corollary \ref{cor: arma_is_concentrated} and
			 Theorem \ref{thm: concent_Rs},
			we know that a D/C-ARMA process with finite discrete space is a typical source, and 
			\ea{
				R^*(\ep)=R_B(\ep)=R(\ep)=d_B\big(\{\Zv_t\}\big)=d(\xi_1)=\alpha,
			}
			where the last equality holds because of Theorem \ref{thm:disc_cont_dim}.
	\end{IEEEproof}

\end{document}